\documentclass{easychair}
\usepackage{doc}
\usepackage{booktabs}
\usepackage{multirow}
\usepackage{booktabs}
\usepackage{multirow}
\usepackage{tabularx}
\usepackage{paralist}
\usepackage[inline]{enumitem} 
\RequirePackage{tikz} 
\tikzset{%
  point/.style={fill=#1, inner sep=1.4pt, circle},
  point/.default=black}
\usetikzlibrary{decorations.pathreplacing, calligraphy, arrows.meta}
\RequirePackage{subcaption} 
\RequirePackage{tabularray} 

\usepackage[inline]{enumitem}
\usepackage{amssymb}
\usepackage{soul}
\usepackage{subcaption}
\usepackage{siunitx}

\newtheorem{theorem}{Theorem}

\newcommand{\RR}{\mathbb{R}} 

\newcommand{\Mat}[3]{
  {#1}^{#2\times#3}}

\newcommand{\Perms}[2][\relax]{
  \Pi_{#2}\ifx#1\relax\relax\else(#1)\fi}
\newcommand{\CV}[2]{
  {#1}^{#2}}

\newcommand{\eg}{e.\,g.}
\newcommand{\ie}{i.\,e.}

\usetikzlibrary{patterns,patterns.meta}

\usepackage{xcolor}
\definecolor{yescolour}{RGB}{127,201,127}
\definecolor{nocolour}{RGB}{190,174,212}
\definecolor{basecolour}{RGB}{91,137,48}
\tikzset{%
  baseline/.style={
    draw=basecolour,
    densely dashed,
    line width= 1.2pt,
    line join=round
  },
  yesstyle/.style={%
    postaction={draw,black!80},
    pattern={crosshatch dots},
    pattern color=yescolour!10!white},
  nostyle/.style={%
    postaction={draw,black!80},
    pattern={crosshatch},
    pattern color=nocolour!30!black}}

\definecolor{palecrimson}{HTML}{C93756}
\definecolor{cinnabar}{HTML}{ff3500}

\title{When Algebraic Symmetry Breaking Meets Solvers:\\ An Experimental Study}

\author{
M\u{a}d\u{a}lina Era\c{s}cu\inst{1}
\and
Johannes Middeke\inst{2}
}
\institute{Faculty of Informatics, West University of Timișoara, Timișoara, Romania\\
  \email{madalina.erascu@e-uvt.ro}
\and
   Temple University, Japan Campus, Tokyo, Japan\\
\email{johannes.middeke@tuj.temple.edu}
}

\authorrunning{Era\c{s}cu, Middeke}
\titlerunning{When Algebraic Symmetry Breaking Meets Solvers: An Experimental Study}

\begin{document}

\maketitle

\begin{abstract}
We present an experimental evaluation of automatically generated polynomial symmetry breaking constraints for integer linear programs. Starting from the method that we introduced at the International Symposium on Symbolic and Algebraic Computation (ISSAC) 2026, we compare solver native quadratic handling, solver-internal reformulation, and explicit linearization on near half-capacity bin-packing benchmarks. 

Experiments with several mathematical programming solvers and satisfiability modulo theory solvers show that the effectiveness of polynomial symmetry breaking is strongly solver-dependent. Compact quadratic breaker families can improve performance, whereas linearization, large breaker sets, or solver reformulations may offset these gains through increased model size or less favorable search behavior. These results suggest that automatically generated symmetry breakers should be evaluated in a solver-aware manner rather than treated as solver-independent additions to a model.
\end{abstract}

\section{Introduction}\label{sec:intro}
Optimization and automated reasoning, including constraint programming~\cite{GentPetriePuget2006SymmetryCP,Walsh2012SymmetryBreakingConstraints}, mixed-integer programming~\cite{Margot2002PruningIsomorphism,kaibel2008packing,salvagnin2018symmetry}, satisfiability~\cite{AloulMarkovSakallahDAC2003Shatter,CrawfordEtAlKR1996}, and satisfiability modulo theory~\cite{DingliwalEtAlArXiv2019SMTSymBreak}, typically treat symmetries as redundant branches of the search tree to be pruned. A common way to address this redundancy is to add constraints which eliminate as many symmetric solutions as possible. These are the so-called symmetry breaking constraints which can be exploited, for example, in a static manner~\cite{puget1993satisfiability,DingliwalEtAlArXiv2019SMTSymBreak}, i.e., adding constraints \emph{a priori} to restrict feasibility to orbit representatives.

In~\cite{erascu2026automaticgenerationpolynomialsymmetry}, we introduced an algebraic method for automatically generating symmetry-breaking constraints. Given the symmetry group of an integer linear program and a polynomial template~$h$, the method constructs breaker families of the form $h(Px)-h(x)\le 0$. Unlike existing constructions, our method naturally yields both linear and nonlinear breakers and can be implemented with standard computer algebra tools. Experiments on bin-packing instances showed that quadratic breakers can outperform both linear variants and Gurobi's \cite{gurobi} built-in symmetry handling.

These results raise a natural question: do the observed gains come from the breakers themselves, or from how a specific solver processes them? This is particularly important for quadratic breakers, since some solvers handle quadratic constraints natively, some reformulate them internally, and others require linear constraints, hence explicit linearization must be applied.

The goal of this paper is to study this solver dependence experimentally with different types of solvers: mathematical programming solvers such as Gurobi \cite{gurobi}, CPLEX \cite{manual1987ibm}, and SCIP \cite{hojny2025scip}, the nonlinear and combinatorial optimization solver Hexaly \cite{hexaly}, and the SMT solver Z3 \cite{de2008z3}.

We make three contributions:
\begin{inparaenum}[(1)]
\item an experimental comparison of native quadratic, solver-reformulated, and explicitly linearized polynomial symmetry breakers;
\item evidence that compact quadratic breakers can help, but their effect is solver- and size-dependent;
\item a reproducible benchmark suite and artifact \cite{github} for evaluating solver-aware symmetry breaking.
\end{inparaenum}

Our experiments show that polynomial symmetry breaking is highly solver-sensitive. Compact quadratic breaker families (few variables, few permutations) often remain beneficial when handled natively, whereas reformulation or linearization may introduce variables and constraints that reduce or eliminate this advantage. Moreover, the same breaker family may improve performance for one solver and slow down another. Thus, automatically generated symmetry breakers should not be evaluated only by their mathematical strength or modelling convenience; their effectiveness must also be assessed together with the preprocessing and search mechanisms of the target solver.
\section{Background: Polynomial Symmetry Breaking}\label{sec:poly-sb}


This section gives an overview over the
symmetry breaker generation which first appeared in~\cite{erascu2026automaticgenerationpolynomialsymmetry}.

We consider \emph{integer linear problems} of the form
\begin{equation}
  \label{ILP}\tag{ILP}
  \text{minimize}\quad f(x),
  \qquad
  \text{subject to}\quad A x \geq b
  \quad\text{and}\quad
  x \in \CV{U}n
\end{equation}
where
\begin{enumerate*}[label={}, itemjoin={\ }, itemjoin*={\ and\ }, afterlabel={}]
\item $f\colon \RR^n \to \RR$ is a function,
\item $x = (x_1, \ldots, x_n)$ is a vector of decision variables,
\item $A \in \Mat{\RR}{m}{n}$ is an $m$-by-$n$ real matrix,
\item $b \in \CV{\RR}{m}$ is a real vector, 
\item $U \subseteq \RR$ is a finite set.
\end{enumerate*}
We will call $f$ the \emph{objective function} of~\eqref{ILP}. The set
$\{ x \in \CV{\RR}{n} \colon A x \geq b \}$ will be refered to as the
\emph{feasible set} of~\eqref{ILP}.

Solvers will typically explore the search space using tree-based
methods such as branch-and-bound. Here, symmetries in the problem
formulation can become problematic since the solver will repeatedly
search essentially identical
subtrees. Following~\cite{erascu2026automaticgenerationpolynomialsymmetry},
we identify symmetries with permutation matrices $P$ where $P$ maps
the feasible set onto itself and fulfils $f(P x) = f(x)$ for all
$x \in \CV{U}n$.

One possible strategy for avoiding the redundant searches that are
caused by the presence of symmetries is to introduce additional
constraints or \emph{(static) symmetry breakers} to the
problem. (See~\cite{erascu2026automaticgenerationpolynomialsymmetry}
for details.) While it is possible to add more than one symmetry
breaker to a problem, one needs to take care that the breakers are
compatible with each other (\ie, that the is at least one optimal
solution which matches all the breakers).

\begin{theorem}[{\cite[Thm.~2.3]{erascu2026automaticgenerationpolynomialsymmetry}}]
  \label{thm} Let $h \in \RR[x]$ be a polynomial with real
  coefficients, and let $G$ be a set of symmetries
  of~\eqref{ILP}. Then the inequalities
  \begin{math}
    \{ h(P x) - h(x) \leq 0 \colon P \in G \}
  \end{math}
  constitute a family of compatible symmetry breakers for~\eqref{ILP}.
\end{theorem}

In order to demonstrate the efficiency of the symmetry breakers
generated through \autoref{thm},
\cite{erascu2026automaticgenerationpolynomialsymmetry} considered a
variation of the bin-packing problem (see, \eg,
\cite[Chapter~8]{martello1990knapsack} for more background on this
problem). We will use the same example which we sketch out here
briefly--see~\cite{erascu2026automaticgenerationpolynomialsymmetry}
for the details. The task is to place $m$ items into as few bins as
possible. For each $k = 1, \ldots, m$, the $k$-th item has size
$s_k > 0$ and each bin can only contain items up to a total size of at
most $B > 0$. The problem can be modelled in the form of~\eqref{ILP}
via
\begin{equation}
  \label{BP}\tag{BP}
  \begin{aligned}
    \text{minimize}\quad
    &
    y_1 + \ldots + y_n \\
    \text{subject to}\quad
    &
    \forall k=1,\ldots,n\colon
    \textstyle\sum_{i=1}^m s_i x_{ik} \leq B y_k,
    \quad
    \forall i=1,\ldots,m\colon
    \textstyle\sum_{k=1}^n x_{ik} = 1, \\
    &
    \forall i = 1,\ldots,m\; 
    \forall k = 1,\ldots,n\colon
    x_{i,k} \in \{0,1\} \text{ and } y_k \in \{0,1\}
  \end{aligned}
\end{equation}
where $y_k = 1$ means that the $k$-th bin is used and $x_{ik} = 1$
means that the $i$-th item is to be placed into the $k$-th~bin.

As in~\cite{erascu2026automaticgenerationpolynomialsymmetry}, we 
generate bin-packing instances with bin size $B = 100$ and item sizes around $B/2 = 50$. We consider 
four different classes: Class~$3$ has three different item sizes ($49, 50, 51$) and similarly classes~$5$, $7$, and~$9$ have five, seven and nine different item sizes nearby $B/2$.

Given an instance of~\eqref{BP}, we use \autoref{thm} to generate a
randomized family of symmetry breakers. For this, we need to know a
set of the symmetries group for which we
use~\cite{erascu2026automaticgenerationpolynomialsymmetry}: Each
symmetry consists of a permutation $\pi$ of the $i$-indices in
$x_{ik}$ such that $s_i = s_{\pi(i)}$ for all $i$ which is combined
with a permutation $\psi$ of the $k$-indices in $x_{ik}$ and
$y_k$. (In other words, $x_{ik}$ maps to $x_{\pi(i)\psi(k)}$ and $y_k$
maps to $y_{\psi(k)}$.)

For our experiments we started with a random polynomial $h \in \RR[x]$
which was generated from a set of fixed templates. These templates
contained homogeneous polynomials of degrees~$1$ or~$2$ as well as
inhomogeneous polynomials of degree~$2$ with a varying total number of
variables. See \autoref{tab:shapes} and \autoref{tab:perm-var-numbers}
in \autoref{app:appendix} for more details. A full table with the
exact parameters for each experiment is included in~\cite{github}. We
then generated random permutation matrices $P_1, \ldots, P_\ell$ where
each permutation was obtained by composing a series of $50$ random $i$-
or $k$-transpositions. From this, we computed the symmetry breaker
family $H = \{ h(P_j x) - h(x) \colon j = 1, \ldots, \ell\}$ where for
the quadratic templates we would additionally filter out any breakers
$h(P_j x) - h(x) \leq 0$ where $\deg (h(P_j x) - h(x)) \leq 1$. (See
the last column ``Keep Only Degree 2'' in \autoref{tab:shapes}.)

\section{Experimental Results}
\label{sec:experimental-results}

We evaluate the performance of automatically generated polynomial symmetry-breaking constraints on randomly generated instances of near half-capacity bin-packing problems. For each benchmark instance, we first solve the original formulation to obtain a Baseline measurement, then generate families of random symmetry breakers using the method from \autoref{sec:poly-sb} and compare the solver performance. All experiments were run on an Apple M2 Pro (10 cores).

\subsection{Gurobi}\label{sec:gurobi}

Gurobi Optimizer \cite{gurobi} (academic
license) was initially used in \cite{erascu2026automaticgenerationpolynomialsymmetry} to test the quadratic symmetry breakers in their original form.  In that experiment, the breakers were passed directly to Gurobi's nonconvex quadratic machinery and compared with the Baseline model and with Gurobi's default settings.  The main outcome was that quadratic breakers consistently outperform linear ones. Moreover, compact (i.e. few variables, few permutations) quadratic breakers, especially \(xy\) templates, were more reliable than linear breakers and often
improved over Gurobi's built-in symmetry handling.  Larger breaker families (i.e. ``many variables'', ``numerous variables'')  slow down the solver because the additional constraints could outweigh the reduction in symmetric search. The detailed experiment is reported in \cite{erascu2026automaticgenerationpolynomialsymmetry}.

For this paper, we ran experiments with $
  n \in \{99,100,700,900\}$.
These values were chosen to align the Gurobi comparison with the range used in
the CPLEX experiments (see \autoref{sec:cplex})\footnote{We initially ran CPLEX with $n \in \{1000, 1024, 2000\}$, as in \cite{erascu2026automaticgenerationpolynomialsymmetry}. However, since the Baseline computation did not complete, we were forced to use smaller values for $n$}. We report
these results separately for the four benchmark classes, because the behaviour
differs substantially by class.

\begin{figure*}[h!]
\centering
\input{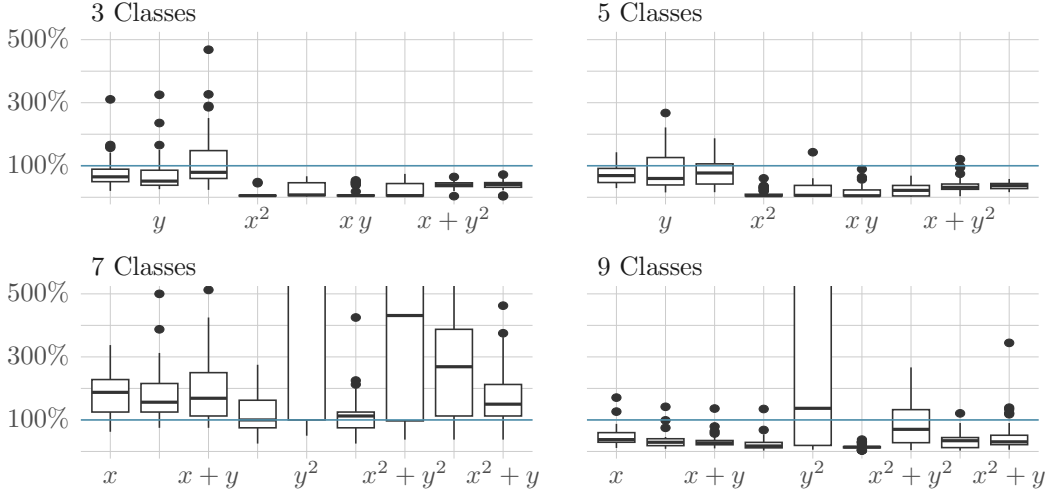} 
\caption{Gurobi relative performance of the tested symmetry-breaking variants (quadratic and linear).
  The \(100\%\) line is the baseline without the generated breakers. Each panel
  is split by benchmark class and by the template used to generate the
  polynomial \(h\).}
\label{fig:gurobi-lpar-tests}
\end{figure*}

\autoref{fig:gurobi-lpar-tests} confirms once more the results from \cite{erascu2026automaticgenerationpolynomialsymmetry}. However,  an exception is observed for class $7$, where performance is generally worse than the Baseline. Since the Baseline is solved in only 0.08 work units, the potential benefit of additional symmetry breaking is limited, while the overhead introduced by the breakers may dominate.

\noindent \textbf{Explicit linearization.}
For comparison with solvers and configurations that handle only linear models,
we tested the following explicit linearization \cite{balas1964extension}: each product of two binary variables is
replaced by an auxiliary binary variable. For instance, for binary variables \(x_1,x_2\), the
product \(x_1x_2\) is represented by a new binary variable \(z_{12}\) and the
constraints
\begin{equation}
  z_{12} - x_1 \leq 0, \qquad
  z_{12} - x_2 \leq 0, \qquad
  x_1 + x_2 - z_{12} \leq 1 .
  \label{eq:binary-product-linearization}
\end{equation}
For example, the quadratic inequality $x_1x_2 + x_3 \leq 1$
is replaced by $z_{12} + x_3 \leq 1$,
together with  constraints in \eqref{eq:binary-product-linearization}.

\begin{figure*}[h!]
\centering
\input{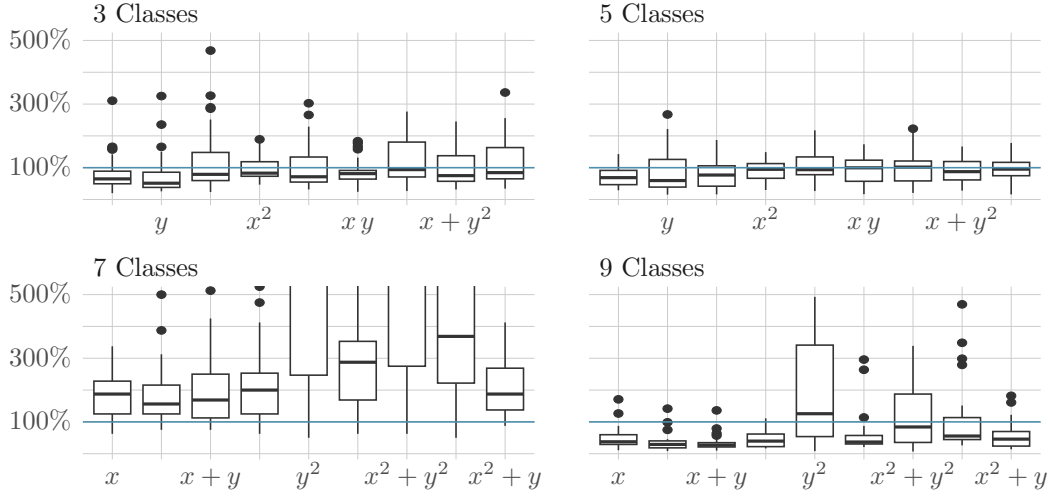}
\caption{Gurobi relative performance of the tested symmetry-breaking variants (liniarized).
  The \(100\%\) line is the Baseline without the generated breakers.  Each panel
  is split by benchmark class and by the template used to generate the
  polynomial \(h\).}
\label{fig:gurobi-linearized}
\end{figure*}

\autoref{fig:gurobi-linearized} shows that linearization changes the performance profile of the breakers. Overall, the quadratic breakers in \autoref{fig:gurobi-lpar-tests} perform better than their linearized version in \autoref{fig:gurobi-linearized}. Linearization generally shows higher values and greater dispersion. For class $7$ this is particularly obvious, but this is because the Baseline was already very fast. Moreover, the linearized quadratic breakers perform similar to the pure linear breakers.

\subsection{CPLEX}
\label{sec:cplex}
CPLEX \cite{manual1987ibm} was first tested on class $3$ instances with \(n=900\), with both
presolve and built-in symmetry handling disabled and with a deterministic time
limit of $4\,000\,000$ ticks.  Under this configuration, linear symmetry
breakers were faster than the Baseline.  For quadratic symmetry breakers, CPLEX
quickly reached solutions very close to optimality, but did not finish within
the deterministic time limit.  The solver logs indicated that CPLEX
reformulated, or internally linearized, the quadratic constraints.  Thus, in
this setting, the solver was not necessarily solving the quadratic constraints
exactly as written in the input model.  We also disabled reductions which limited presolve
reductions, but this did not provide a general mechanism for preserving arbitrary nonconvex quadratic constraints in their original form.  The internal
reformulation can produce strong relaxations that quickly reduce the optimality
gap, but it may also leave the solver with a difficult final search near
optimality.  For this reason, the main CPLEX comparison uses only the explicit
linearizations described above.

\begin{figure*}[h!]
\centering
\input{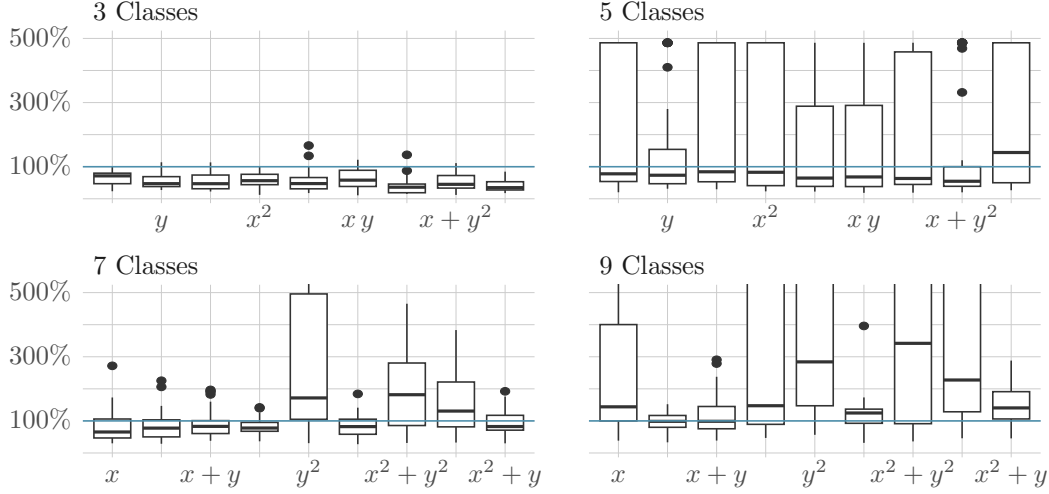}
\caption{CPLEX relative performance of the tested symmetry-breaking variants (liniarized).
  The \(100\%\) line is the Baseline without the generated breakers. Each panel
  is split by benchmark class and by the template used to generate the
  polynomial \(h\).}
\label{fig:CPLEX-liniarized}
\end{figure*}

\autoref{fig:CPLEX-liniarized} shows a different pattern from the Gurobi
linearization in \autoref{fig:gurobi-linearized}.  For class $3$,
CPLEX is highly stable: almost all templates are below the Baseline and the
spread is small. Classes $5$ and $9$ are much less stable.  In both panels, several
templates have large upper ranges or outliers, showing that the same
linearized breakers can produce substantial slowdowns on other benchmark
classes. Class $7$ is mixed: the simpler templates are close to the Baseline or
slightly below it, but \(y^2\) and larger combined templates can increase solver effort.
\subsection{SCIP and Hexaly}
\label{subsec:scip-hexaly-results}

We evaluated the SCIP Optimization Suite \cite{hojny2025scip} under two solver configurations: the default configuration, using SCIP's default presolve and symmetry-handling settings, and a second configuration in which both presolve and symmetry handling were disabled. For each configuration, we compared the Baseline formulation with the ``few variables, few permutations, linearized $x$'' variant. The results are reported in \autoref{tab:scip-default-disabled} in \autoref{app:appendix}; for large $n$, $n \in \{700, 900\}$, neither configuration solved any instance within the $1800$ seconds CPU-time limit. For the small $n$, $n \in \{99, 100\}$, the instances were solved, but the running-time effect of the symmetry breakers was not consistent.

We also considered Hexaly (academic license) \cite{hexaly}, a solver for nonlinear and combinatorial optimization models. Unlike classical mixed-integer linear programming solvers, Hexaly represents models through expression graphs and can operate directly on nonlinear expressions, constraints, and objectives. Nevertheless, we did not investigate static symmetry-breaking constraints further in Hexaly; based on the available documentation and communication with the developers, such constraints may interact unfavorably with the solver's internal search mechanisms, making their effect difficult to control or predict in this setting.

\subsection{Z3}
\label{sec:z3}

We tested Z3 optimization feature \cite{10.1007/978-3-662-46681-0_14} on smaller instances, using \(n=99\) and class \(9\) (motivated by the CPLEX instance), with
a CPU-time limit of 10 minutes. 
The Baseline optimization problem did not end within the time limit, nor the problems accompanied by symmetry breakers.

We further considered satisfiability-only problems obtained by dropping the
objective function, in order to analyze the effect of the
symmetry-breaking constraints.

\begin{figure*}[h!]
\centering
\input{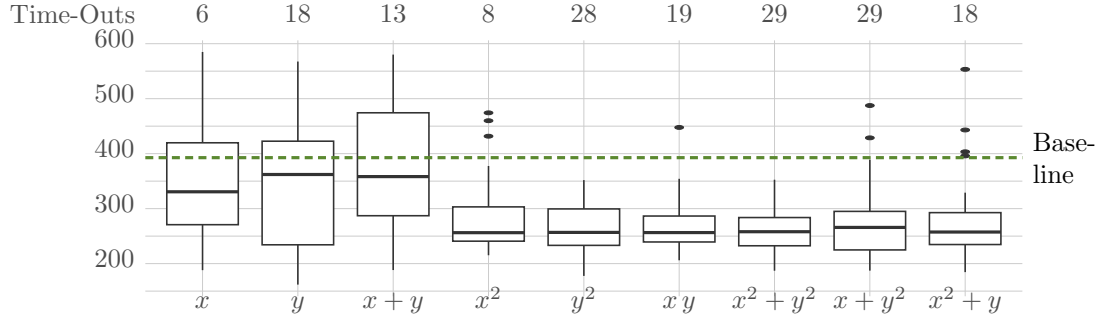}
\caption{Timings for the Z3 experiments (tactic \texttt{smt} enabled) grouped by the template shape. The times are shown in seconds. The graph considers only the instances which did finish within the time bound of \SI{10}{\minute}.}
\label{fig:z3omt-time-per-shape_tactic_smt}
\end{figure*}

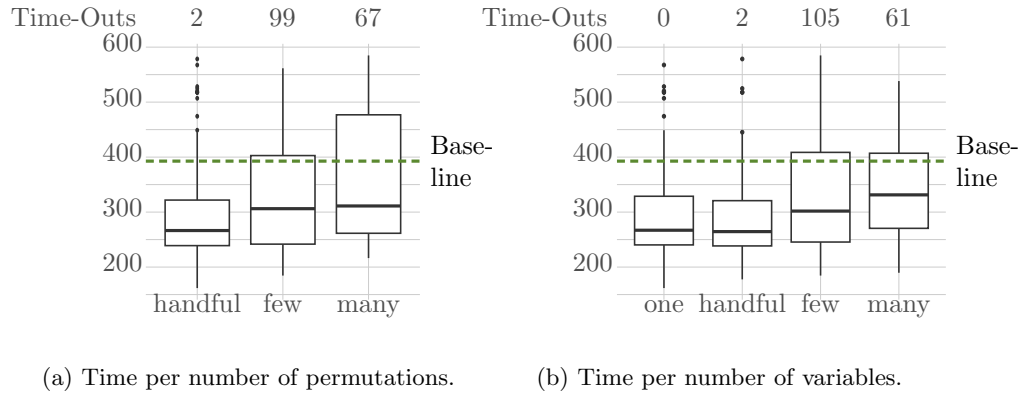
\begin{figure*}[h!]
  \centering
  \begin{subfigure}[t]{0.45\textwidth}
    \centering
\begin{tikzpicture}[x=0.25pt,y=0.35pt]
\definecolor{fillColor}{RGB}{255,255,255}
\begin{scope}
\definecolor{fillColor}{RGB}{255,255,255}

\end{scope}
\begin{scope}
  \definecolor{drawColor}{gray}{0.30}
  
  \node[left, text=drawColor] at (63, 360) {Time-Outs};
  \node[text=drawColor] at (140, 360) {2};
  \node[text=drawColor] at (269, 360) {99};
  \node[text=drawColor] at (398, 360) {67};
\end{scope}
\begin{scope}
\definecolor{drawColor}{gray}{0.80}

\path[draw=drawColor,line width= 0.2pt,line join=round] ( 63.24, 61.67) --
	(475.89, 61.67);

\path[draw=drawColor,line width= 0.2pt,line join=round] ( 63.24,120.80) --
	(475.89,120.80);

\path[draw=drawColor,line width= 0.2pt,line join=round] ( 63.24,179.94) --
	(475.89,179.94);

\path[draw=drawColor,line width= 0.2pt,line join=round] ( 63.24,239.07) --
	(475.89,239.07);

\path[draw=drawColor,line width= 0.2pt,line join=round] ( 63.24,298.20) --
	(475.89,298.20);

\path[draw=drawColor,line width= 0.2pt,line join=round] ( 63.24, 91.24) --
	(475.89, 91.24);

\path[draw=drawColor,line width= 0.2pt,line join=round] ( 63.24,150.37) --
	(475.89,150.37);

\path[draw=drawColor,line width= 0.2pt,line join=round] ( 63.24,209.50) --
	(475.89,209.50);

\path[draw=drawColor,line width= 0.2pt,line join=round] ( 63.24,268.63) --
	(475.89,268.63);

\path[draw=drawColor,line width= 0.2pt,line join=round] ( 63.24,327.76) --
	(475.89,327.76);

\path[draw=drawColor,line width= 0.2pt,line join=round] (140.61, 56.15) --
	(140.61,331.35);

\path[draw=drawColor,line width= 0.2pt,line join=round] (269.57, 56.15) --
	(269.57,331.35);

\path[draw=drawColor,line width= 0.2pt,line join=round] (398.52, 56.15) --
	(398.52,331.35);
\definecolor{drawColor}{gray}{0.20}
\definecolor{fillColor}{gray}{0.20}

\path[draw=drawColor,line width= 0.4pt,line join=round,line cap=round,fill=fillColor] (140.61,279.26) circle (  2.04);

\path[draw=drawColor,line width= 0.4pt,line join=round,line cap=round,fill=fillColor] (140.61,279.06) circle (  2.04);

\path[draw=drawColor,line width= 0.4pt,line join=round,line cap=round,fill=fillColor] (140.61,315.04) circle (  2.04);

\path[draw=drawColor,line width= 0.4pt,line join=round,line cap=round,fill=fillColor] (140.61,283.28) circle (  2.04);

\path[draw=drawColor,line width= 0.4pt,line join=round,line cap=round,fill=fillColor] (140.61,238.42) circle (  2.04);

\path[draw=drawColor,line width= 0.4pt,line join=round,line cap=round,fill=fillColor] (140.61,278.93) circle (  2.04);

\path[draw=drawColor,line width= 0.4pt,line join=round,line cap=round,fill=fillColor] (140.61,285.34) circle (  2.04);

\path[draw=drawColor,line width= 0.4pt,line join=round,line cap=round,fill=fillColor] (140.61,280.64) circle (  2.04);

\path[draw=drawColor,line width= 0.4pt,line join=round,line cap=round,fill=fillColor] (140.61,308.55) circle (  2.04);

\path[draw=drawColor,line width= 0.4pt,line join=round,line cap=round,fill=fillColor] (140.61,272.77) circle (  2.04);

\path[draw=drawColor,line width= 0.4pt,line join=round,line cap=round,fill=fillColor] (140.61,279.71) circle (  2.04);

\path[draw=drawColor,line width= 0.4pt,line join=round,line cap=round,fill=fillColor] (140.61,253.44) circle (  2.04);

\path[draw=drawColor,line width= 0.6pt,line join=round] (140.61,163.30) -- (140.61,236.29);

\path[draw=drawColor,line width= 0.6pt,line join=round] (140.61,114.31) -- (140.61, 68.66);
\definecolor{fillColor}{RGB}{255,255,255}

\path[draw=drawColor,line width= 0.6pt,fill=fillColor] ( 92.26,163.30) --
	( 92.26,114.31) --
	(188.97,114.31) --
	(188.97,163.30) --
	( 92.26,163.30) --
	cycle;

\path[draw=drawColor,line width= 1.2pt] ( 92.26,130.53) -- (188.97,130.53);

\path[draw=drawColor,line width= 0.6pt,line join=round] (269.57,211.15) -- (269.57,305.04);

\path[draw=drawColor,line width= 0.6pt,line join=round] (269.57,115.91) -- (269.57, 82.10);

\path[draw=drawColor,line width= 0.6pt,fill=fillColor] (221.21,211.15) --
	(221.21,115.91) --
	(317.92,115.91) --
	(317.92,211.15) --
	(221.21,211.15) --
	cycle;

\path[draw=drawColor,line width= 1.2pt] (221.21,154.04) -- (317.92,154.04);

\path[draw=drawColor,line width= 0.6pt,line join=round] (398.52,254.99) -- (398.52,318.84);

\path[draw=drawColor,line width= 0.6pt,line join=round] (398.52,127.59) -- (398.52,100.92);

\path[draw=drawColor,line width= 0.6pt,fill=fillColor] (350.16,254.99) --
	(350.16,127.59) --
	(446.88,127.59) --
	(446.88,254.99) --
	(350.16,254.99) --
	cycle;

\path[draw=drawColor,line width= 1.2pt] (350.16,156.96) -- (446.88,156.96);
\definecolor{drawColor}{RGB}{255,0,0}

\path[baseline] ( 63.24,205.18) -- (475.89,205.18)
node[right, align=left] {Base-\\line};

\end{scope}
\begin{scope}
\definecolor{drawColor}{gray}{0.30}

\node[text=drawColor,anchor=base east,inner sep=0pt, outer sep=0pt] at ( 58.07, 87.28) {200};

\node[text=drawColor,anchor=base east,inner sep=0pt, outer sep=0pt] at ( 58.07,146.41) {300};

\node[text=drawColor,anchor=base east,inner sep=0pt, outer sep=0pt] at ( 58.07,205.54) {400};

\node[text=drawColor,anchor=base east,inner sep=0pt, outer sep=0pt] at ( 58.07,264.67) {500};

\node[text=drawColor,anchor=base east,inner sep=0pt, outer sep=0pt] at ( 58.07,323.80) {600};
\end{scope}
\begin{scope}
\definecolor{drawColor}{gray}{0.30}

\node[text=drawColor,anchor=base,inner sep=0pt, outer sep=0pt] at (140.61, 43.06) {handful};

\node[text=drawColor,anchor=base,inner sep=0pt, outer sep=0pt] at (269.57, 43.06) {few};

\node[text=drawColor,anchor=base,inner sep=0pt, outer sep=0pt] at (398.52, 43.06) {many};
\end{scope}
\end{tikzpicture}
    \caption{Time per number of permutations.}
    \label{fig:z3omt-time-per-perms-tactic-smt}
  \end{subfigure}
  \hspace{-1.5em}
  \begin{subfigure}[t]{0.45\textwidth}
    \centering
\begin{tikzpicture}[x=0.3pt,y=0.35pt]
\definecolor{fillColor}{RGB}{255,255,255}
\begin{scope}
\definecolor{fillColor}{RGB}{255,255,255}

\end{scope}
\begin{scope}
  \definecolor{drawColor}{gray}{0.30}
  
  \node[left, text=drawColor] at (63, 360) {Time-Outs};

  \node[text=drawColor] at (122.19,360) {0};
  \node[text=drawColor] at (220.44,360) {2};
  \node[text=drawColor] at (318.69,360) {105};
  \node[text=drawColor] at (416.94,360) {61};
\end{scope}
\begin{scope}
\definecolor{drawColor}{gray}{0.80}

\path[draw=drawColor,line width= 0.2pt,line join=round] ( 63.24, 61.67) --
	(475.89, 61.67);

\path[draw=drawColor,line width= 0.2pt,line join=round] ( 63.24,120.80) --
	(475.89,120.80);

\path[draw=drawColor,line width= 0.2pt,line join=round] ( 63.24,179.94) --
	(475.89,179.94);

\path[draw=drawColor,line width= 0.2pt,line join=round] ( 63.24,239.07) --
	(475.89,239.07);

\path[draw=drawColor,line width= 0.2pt,line join=round] ( 63.24,298.20) --
	(475.89,298.20);

\path[draw=drawColor,line width= 0.2pt,line join=round] ( 63.24, 91.24) --
	(475.89, 91.24);

\path[draw=drawColor,line width= 0.2pt,line join=round] ( 63.24,150.37) --
	(475.89,150.37);

\path[draw=drawColor,line width= 0.2pt,line join=round] ( 63.24,209.50) --
	(475.89,209.50);

\path[draw=drawColor,line width= 0.2pt,line join=round] ( 63.24,268.63) --
	(475.89,268.63);

\path[draw=drawColor,line width= 0.2pt,line join=round] ( 63.24,327.76) --
	(475.89,327.76);

\path[draw=drawColor,line width= 0.2pt,line join=round] (122.19, 56.15) --
	(122.19,331.35);

\path[draw=drawColor,line width= 0.2pt,line join=round] (220.44, 56.15) --
	(220.44,331.35);

\path[draw=drawColor,line width= 0.2pt,line join=round] (318.69, 56.15) --
	(318.69,331.35);

\path[draw=drawColor,line width= 0.2pt,line join=round] (416.94, 56.15) --
	(416.94,331.35);
\definecolor{drawColor}{gray}{0.20}
\definecolor{fillColor}{gray}{0.20}

\path[draw=drawColor,line width= 0.4pt,line join=round,line cap=round,fill=fillColor] (122.19,278.93) circle (  2.04);

\path[draw=drawColor,line width= 0.4pt,line join=round,line cap=round,fill=fillColor] (122.19,285.34) circle (  2.04);

\path[draw=drawColor,line width= 0.4pt,line join=round,line cap=round,fill=fillColor] (122.19,280.64) circle (  2.04);

\path[draw=drawColor,line width= 0.4pt,line join=round,line cap=round,fill=fillColor] (122.19,308.55) circle (  2.04);

\path[draw=drawColor,line width= 0.4pt,line join=round,line cap=round,fill=fillColor] (122.19,272.77) circle (  2.04);

\path[draw=drawColor,line width= 0.4pt,line join=round,line cap=round,fill=fillColor] (122.19,279.71) circle (  2.04);

\path[draw=drawColor,line width= 0.4pt,line join=round,line cap=round,fill=fillColor] (122.19,253.44) circle (  2.04);

\path[draw=drawColor,line width= 0.6pt,line join=round] (122.19,167.40) -- (122.19,238.42);

\path[draw=drawColor,line width= 0.6pt,line join=round] (122.19,115.08) -- (122.19, 68.66);
\definecolor{fillColor}{RGB}{255,255,255}

\path[draw=drawColor,line width= 0.6pt,fill=fillColor] ( 85.35,167.40) --
	( 85.35,115.08) --
	(159.04,115.08) --
	(159.04,167.40) --
	( 85.35,167.40) --
	cycle;

\path[draw=drawColor,line width= 1.2pt] ( 85.35,130.87) -- (159.04,130.87);
\definecolor{fillColor}{gray}{0.20}

\path[draw=drawColor,line width= 0.4pt,line join=round,line cap=round,fill=fillColor] (220.44,279.26) circle (  2.04);

\path[draw=drawColor,line width= 0.4pt,line join=round,line cap=round,fill=fillColor] (220.44,279.06) circle (  2.04);

\path[draw=drawColor,line width= 0.4pt,line join=round,line cap=round,fill=fillColor] (220.44,315.04) circle (  2.04);

\path[draw=drawColor,line width= 0.4pt,line join=round,line cap=round,fill=fillColor] (220.44,283.28) circle (  2.04);

\path[draw=drawColor,line width= 0.4pt,line join=round,line cap=round,fill=fillColor] (220.44,236.29) circle (  2.04);

\path[draw=drawColor,line width= 0.6pt,line join=round] (220.44,162.66) -- (220.44,234.94);

\path[draw=drawColor,line width= 0.6pt,line join=round] (220.44,113.98) -- (220.44, 78.02);
\definecolor{fillColor}{RGB}{255,255,255}

\path[draw=drawColor,line width= 0.6pt,fill=fillColor] (183.60,162.66) --
	(183.60,113.98) --
	(257.29,113.98) --
	(257.29,162.66) --
	(183.60,162.66) --
	cycle;

\path[draw=drawColor,line width= 1.2pt] (183.60,129.38) -- (257.29,129.38);

\path[draw=drawColor,line width= 0.6pt,line join=round] (318.69,214.59) -- (318.69,318.84);

\path[draw=drawColor,line width= 0.6pt,line join=round] (318.69,118.17) -- (318.69, 82.10);

\path[draw=drawColor,line width= 0.6pt,fill=fillColor] (281.85,214.59) --
	(281.85,118.17) --
	(355.53,118.17) --
	(355.53,214.59) --
	(281.85,214.59) --
	cycle;

\path[draw=drawColor,line width= 1.2pt] (281.85,151.45) -- (355.53,151.45);

\path[draw=drawColor,line width= 0.6pt,line join=round] (416.94,213.66) -- (416.94,291.29);

\path[draw=drawColor,line width= 0.6pt,line join=round] (416.94,132.90) -- (416.94, 85.12);

\path[draw=drawColor,line width= 0.6pt,fill=fillColor] (380.10,213.66) --
	(380.10,132.90) --
	(453.78,132.90) --
	(453.78,213.66) --
	(380.10,213.66) --
	cycle;

\path[draw=drawColor,line width= 1.2pt] (380.10,168.90) -- (453.78,168.90);
\definecolor{drawColor}{RGB}{255,0,0}

\path[baseline] ( 63.24,205.18) -- (475.89,205.18)
node[right, align=left] {Base-\\line};
\end{scope}
\begin{scope}
\definecolor{drawColor}{gray}{0.30}

\node[text=drawColor,anchor=base east,inner sep=0pt, outer sep=0pt] at ( 58.07, 87.28) {200};

\node[text=drawColor,anchor=base east,inner sep=0pt, outer sep=0pt] at ( 58.07,146.41) {300};

\node[text=drawColor,anchor=base east,inner sep=0pt, outer sep=0pt] at ( 58.07,205.54) {400};

\node[text=drawColor,anchor=base east,inner sep=0pt, outer sep=0pt] at ( 58.07,264.67) {500};

\node[text=drawColor,anchor=base east,inner sep=0pt, outer sep=0pt] at ( 58.07,323.80) {600};
\end{scope}
\begin{scope}
\definecolor{drawColor}{gray}{0.30}

\node[text=drawColor,anchor=base,inner sep=0pt, outer sep=0pt] at (122.19, 43.06) {one};

\node[text=drawColor,anchor=base,inner sep=0pt, outer sep=0pt] at (220.44, 43.06) {handful};

\node[text=drawColor,anchor=base,inner sep=0pt, outer sep=0pt] at (318.69, 43.06) {few};

\node[text=drawColor,anchor=base,inner sep=0pt, outer sep=0pt] at (416.94, 43.06) {many};
\end{scope}
\end{tikzpicture}
    \caption{Time per number of variables.}
    \label{fig:z3omt-time-per-vars-tactic-smt}
  \end{subfigure}
  \caption{Timings for Z3 experiments (tactic \texttt{smt} enabled) grouped by number of variables
or number of permutations. The times are shown in seconds. The graph considers only the instances which did finish within the time bound of \SI{10}{\minute}.}
\label{fig:z3omt-time-per-shape-tactic-smt}
\end{figure*}

\autoref{fig:z3omt-time-per-shape_tactic_smt} and \autoref{fig:z3omt-time-per-shape-tactic-smt} summarize the satisfiability-only Z3 experiments with the tactic \texttt{smt} enabled. In this configuration, Z3 uses its standard SMT engine. The dashed line represents the Baseline. \autoref{fig:z3omt-time-per-shape_tactic_smt} confirms that the generated breakers can help Z3. However, the solved-time distributions must be interpreted together with the time-out counts, with $x$ and $x^2$ instances providing the best balance between robustness and runtime.  \autoref{fig:z3omt-time-per-shape-tactic-smt} shows that breaker size is also important with the combination ``handful'' for permutations and ``one'' or ``handful'' for variables being the most efficient ones. 

\autoref{fig:z3-time-per-shape_no_tactic_smt} and \autoref{fig:z3-time-per-vars-perms_no_tactic_smt}  summarize the satisfiability-only Z3 experiments with the tactic \texttt{smt} disabled. In this configuration, the logs revealed that Z3 uses a pseudo-Boolean/bit-vector-backed engine. Note that the Baseline case did not finish in the allotted time, so it is not pictured in the figures. 
\autoref{fig:z3-time-per-vars-perms_no_tactic_smt} suggests that the number of variables and permutations used in the quadratic symmetry breakers must be balanced carefully, with "handful", "handful" being the best combination. This indicates that effective symmetry breaking depends not only on the encoding template, but also on choosing a suitable breaker size. 
 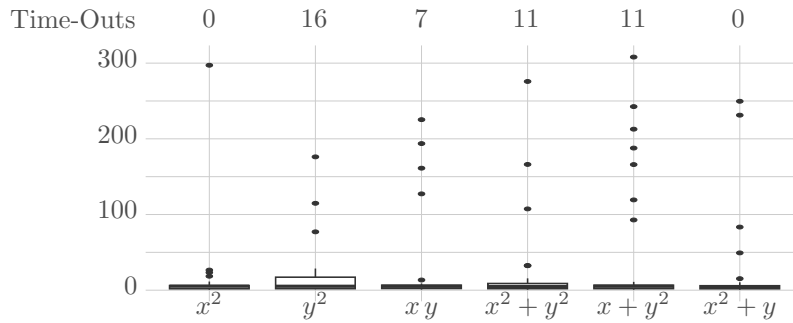
\begin{figure*}[h!]
\centering
\begin{tikzpicture}[x=0.6pt,y=0.35pt]
\definecolor{fillColor}{RGB}{255,255,255}
\begin{scope}
\definecolor{fillColor}{RGB}{255,255,255}

\end{scope}
\begin{scope}
  \definecolor{drawColor}{gray}{0.30}
  
  \node[left, text=drawColor] at (63, 360) {Time-Outs};
  \node[text=drawColor] at (103, 360) {0};
  \node[text=drawColor] at (169, 360) {16};
  \node[text=drawColor] at (236, 360) {7};
  \node[text=drawColor] at (302, 360) {11};
  \node[text=drawColor] at (369, 360) {11};
  \node[text=drawColor] at (435, 360) {0};
\end{scope}
\begin{scope}
\definecolor{drawColor}{gray}{0.80}

\path[draw=drawColor,line width= 0.2pt,line join=round] ( 63.24,108.83) --
	(475.89,108.83);

\path[draw=drawColor,line width= 0.2pt,line join=round] ( 63.24,190.22) --
	(475.89,190.22);

\path[draw=drawColor,line width= 0.2pt,line join=round] ( 63.24,271.61) --
	(475.89,271.61);

\path[draw=drawColor,line width= 0.2pt,line join=round] ( 63.24, 68.13) --
	(475.89, 68.13);

\path[draw=drawColor,line width= 0.2pt,line join=round] ( 63.24,149.52) --
	(475.89,149.52);

\path[draw=drawColor,line width= 0.2pt,line join=round] ( 63.24,230.91) --
	(475.89,230.91);

\path[draw=drawColor,line width= 0.2pt,line join=round] ( 63.24,312.31) --
	(475.89,312.31);

\path[draw=drawColor,line width= 0.2pt,line join=round] (103.18, 56.15) --
	(103.18,331.35);

\path[draw=drawColor,line width= 0.2pt,line join=round] (169.73, 56.15) --
	(169.73,331.35);

\path[draw=drawColor,line width= 0.2pt,line join=round] (236.29, 56.15) --
	(236.29,331.35);

\path[draw=drawColor,line width= 0.2pt,line join=round] (302.84, 56.15) --
	(302.84,331.35);

\path[draw=drawColor,line width= 0.2pt,line join=round] (369.40, 56.15) --
	(369.40,331.35);

\path[draw=drawColor,line width= 0.2pt,line join=round] (435.96, 56.15) --
	(435.96,331.35);
\definecolor{drawColor}{gray}{0.20}
\definecolor{fillColor}{gray}{0.20}

\path[draw=drawColor,line width= 0.4pt,line join=round,line cap=round,fill=fillColor] (103.18, 89.77) circle (  2.04);

\path[draw=drawColor,line width= 0.4pt,line join=round,line cap=round,fill=fillColor] (103.18, 83.15) circle (  2.04);

\path[draw=drawColor,line width= 0.4pt,line join=round,line cap=round,fill=fillColor] (103.18, 87.17) circle (  2.04);

\path[draw=drawColor,line width= 0.4pt,line join=round,line cap=round,fill=fillColor] (103.18,310.05) circle (  2.04);

\path[draw=drawColor,line width= 0.6pt,line join=round] (103.18, 73.13) -- (103.18, 77.44);

\path[draw=drawColor,line width= 0.6pt,line join=round] (103.18, 69.67) -- (103.18, 68.66);
\definecolor{fillColor}{RGB}{255,255,255}

\path[draw=drawColor,line width= 0.6pt,fill=fillColor] ( 78.22, 73.13) --
	( 78.22, 69.67) --
	(128.14, 69.67) --
	(128.14, 73.13) --
	( 78.22, 73.13) --
	cycle;

\path[draw=drawColor,line width= 1.2pt] ( 78.22, 72.44) -- (128.14, 72.44);
\definecolor{fillColor}{gray}{0.20}

\path[draw=drawColor,line width= 0.4pt,line join=round,line cap=round,fill=fillColor] (169.73,130.89) circle (  2.04);

\path[draw=drawColor,line width= 0.4pt,line join=round,line cap=round,fill=fillColor] (169.73,161.58) circle (  2.04);

\path[draw=drawColor,line width= 0.4pt,line join=round,line cap=round,fill=fillColor] (169.73,211.52) circle (  2.04);

\path[draw=drawColor,line width= 0.6pt,line join=round] (169.73, 82.12) -- (169.73, 91.42);

\path[draw=drawColor,line width= 0.6pt,line join=round] (169.73, 69.77) -- (169.73, 69.25);
\definecolor{fillColor}{RGB}{255,255,255}

\path[draw=drawColor,line width= 0.6pt,fill=fillColor] (144.77, 82.12) --
	(144.77, 69.77) --
	(194.69, 69.77) --
	(194.69, 82.12) --
	(144.77, 82.12) --
	cycle;

\path[draw=drawColor,line width= 1.2pt] (144.77, 72.38) -- (194.69, 72.38);
\definecolor{fillColor}{gray}{0.20}

\path[draw=drawColor,line width= 0.4pt,line join=round,line cap=round,fill=fillColor] (236.29,225.81) circle (  2.04);

\path[draw=drawColor,line width= 0.4pt,line join=round,line cap=round,fill=fillColor] (236.29,171.73) circle (  2.04);

\path[draw=drawColor,line width= 0.4pt,line join=round,line cap=round,fill=fillColor] (236.29,199.34) circle (  2.04);

\path[draw=drawColor,line width= 0.4pt,line join=round,line cap=round,fill=fillColor] (236.29,251.52) circle (  2.04);

\path[draw=drawColor,line width= 0.4pt,line join=round,line cap=round,fill=fillColor] (236.29, 79.09) circle (  2.04);

\path[draw=drawColor,line width= 0.6pt,line join=round] (236.29, 73.55) -- (236.29, 75.36);

\path[draw=drawColor,line width= 0.6pt,line join=round] (236.29, 69.96) -- (236.29, 68.85);
\definecolor{fillColor}{RGB}{255,255,255}

\path[draw=drawColor,line width= 0.6pt,fill=fillColor] (211.33, 73.55) --
	(211.33, 69.96) --
	(261.25, 69.96) --
	(261.25, 73.55) --
	(211.33, 73.55) --
	cycle;

\path[draw=drawColor,line width= 1.2pt] (211.33, 72.29) -- (261.25, 72.29);
\definecolor{fillColor}{gray}{0.20}

\path[draw=drawColor,line width= 0.4pt,line join=round,line cap=round,fill=fillColor] (302.84,155.56) circle (  2.04);

\path[draw=drawColor,line width= 0.4pt,line join=round,line cap=round,fill=fillColor] (302.84,292.61) circle (  2.04);

\path[draw=drawColor,line width= 0.4pt,line join=round,line cap=round,fill=fillColor] (302.84, 94.44) circle (  2.04);

\path[draw=drawColor,line width= 0.4pt,line join=round,line cap=round,fill=fillColor] (302.84, 94.68) circle (  2.04);

\path[draw=drawColor,line width= 0.4pt,line join=round,line cap=round,fill=fillColor] (302.84,203.39) circle (  2.04);

\path[draw=drawColor,line width= 0.6pt,line join=round] (302.84, 75.35) -- (302.84, 80.86);

\path[draw=drawColor,line width= 0.6pt,line join=round] (302.84, 69.78) -- (302.84, 68.69);
\definecolor{fillColor}{RGB}{255,255,255}

\path[draw=drawColor,line width= 0.6pt,fill=fillColor] (277.89, 75.35) --
	(277.89, 69.78) --
	(327.80, 69.78) --
	(327.80, 75.35) --
	(277.89, 75.35) --
	cycle;

\path[draw=drawColor,line width= 1.2pt] (277.89, 72.07) -- (327.80, 72.07);
\definecolor{fillColor}{gray}{0.20}

\path[draw=drawColor,line width= 0.4pt,line join=round,line cap=round,fill=fillColor] (369.40,203.23) circle (  2.04);

\path[draw=drawColor,line width= 0.4pt,line join=round,line cap=round,fill=fillColor] (369.40,220.96) circle (  2.04);

\path[draw=drawColor,line width= 0.4pt,line join=round,line cap=round,fill=fillColor] (369.40,318.84) circle (  2.04);

\path[draw=drawColor,line width= 0.4pt,line join=round,line cap=round,fill=fillColor] (369.40,241.29) circle (  2.04);

\path[draw=drawColor,line width= 0.4pt,line join=round,line cap=round,fill=fillColor] (369.40,143.66) circle (  2.04);

\path[draw=drawColor,line width= 0.4pt,line join=round,line cap=round,fill=fillColor] (369.40,265.57) circle (  2.04);

\path[draw=drawColor,line width= 0.4pt,line join=round,line cap=round,fill=fillColor] (369.40,165.23) circle (  2.04);

\path[draw=drawColor,line width= 0.6pt,line join=round] (369.40, 73.51) -- (369.40, 77.08);

\path[draw=drawColor,line width= 0.6pt,line join=round] (369.40, 69.80) -- (369.40, 68.69);
\definecolor{fillColor}{RGB}{255,255,255}

\path[draw=drawColor,line width= 0.6pt,fill=fillColor] (344.44, 73.51) --
	(344.44, 69.80) --
	(394.36, 69.80) --
	(394.36, 73.51) --
	(344.44, 73.51) --
	cycle;

\path[draw=drawColor,line width= 1.2pt] (344.44, 72.35) -- (394.36, 72.35);
\definecolor{fillColor}{gray}{0.20}

\path[draw=drawColor,line width= 0.4pt,line join=round,line cap=round,fill=fillColor] (435.96,271.30) circle (  2.04);

\path[draw=drawColor,line width= 0.4pt,line join=round,line cap=round,fill=fillColor] (435.96,108.24) circle (  2.04);

\path[draw=drawColor,line width= 0.4pt,line join=round,line cap=round,fill=fillColor] (435.96,256.35) circle (  2.04);

\path[draw=drawColor,line width= 0.4pt,line join=round,line cap=round,fill=fillColor] (435.96, 80.54) circle (  2.04);

\path[draw=drawColor,line width= 0.4pt,line join=round,line cap=round,fill=fillColor] (435.96,136.02) circle (  2.04);

\path[draw=drawColor,line width= 0.6pt,line join=round] (435.96, 72.94) -- (435.96, 76.64);

\path[draw=drawColor,line width= 0.6pt,line join=round] (435.96, 69.65) -- (435.96, 68.66);
\definecolor{fillColor}{RGB}{255,255,255}

\path[draw=drawColor,line width= 0.6pt,fill=fillColor] (411.00, 72.94) --
	(411.00, 69.65) --
	(460.91, 69.65) --
	(460.91, 72.94) --
	(411.00, 72.94) --
	cycle;

\path[draw=drawColor,line width= 1.2pt] (411.00, 71.86) -- (460.91, 71.86);
\end{scope}
\begin{scope}
\definecolor{drawColor}{gray}{0.30}

\node[text=drawColor,anchor=base east,inner sep=0pt, outer sep=0pt] at ( 58.07, 64.17) {0};

\node[text=drawColor,anchor=base east,inner sep=0pt, outer sep=0pt] at ( 58.07,145.56) {100};

\node[text=drawColor,anchor=base east,inner sep=0pt, outer sep=0pt] at ( 58.07,226.95) {200};

\node[text=drawColor,anchor=base east,inner sep=0pt, outer sep=0pt] at ( 58.07,308.35) {300};
\end{scope}
\begin{scope}
\definecolor{drawColor}{gray}{0.30}

\node[text=drawColor,anchor=base,inner sep=0pt, outer sep=0pt] at (103.18, 43.06) {$x^2$};

\node[text=drawColor,anchor=base,inner sep=0pt, outer sep=0pt] at (169.73, 43.06) {$y^2$};

\node[text=drawColor,anchor=base,inner sep=0pt, outer sep=0pt] at (236.29, 43.06) {$x\,y$};

\node[text=drawColor,anchor=base,inner sep=0pt, outer sep=0pt] at (302.84, 43.06) {$x^2 + y^2$};

\node[text=drawColor,anchor=base,inner sep=0pt, outer sep=0pt] at (369.40, 43.06) {$x + y^2$};

\node[text=drawColor,anchor=base,inner sep=0pt, outer sep=0pt] at (435.96, 43.06) {$x^2 + y$};
\end{scope}
\end{tikzpicture}
\caption{Timings for the Z3 experiments (tactic \texttt{smt} disabled) grouped by the template shape. The times are shown in seconds. The graph considers only the instances which did finish within the time bound of \SI{6}{\minute}. }
\label{fig:z3-time-per-shape_no_tactic_smt}
\end{figure*}
\begin{figure*}[h!]
  \centering
  \begin{subfigure}[t]{0.45\textwidth}
    \centering
\begin{tikzpicture}[x=0.35pt,y=0.35pt]
\definecolor{fillColor}{RGB}{255,255,255}
\begin{scope}
\definecolor{fillColor}{RGB}{255,255,255}

\end{scope}
\begin{scope}
  \definecolor{drawColor}{gray}{0.30}
  
  \node[left, text=drawColor] at (63, 360) {Time-Outs};
  \node[text=drawColor] at (140, 360) {6};
  \node[text=drawColor] at (269, 360) {34};
  \node[text=drawColor] at (398, 360) {5};
\end{scope}
\begin{scope}
\definecolor{drawColor}{gray}{0.80}

\path[draw=drawColor,line width= 0.2pt,line join=round] ( 63.24,108.83) --
	(475.89,108.83);

\path[draw=drawColor,line width= 0.2pt,line join=round] ( 63.24,190.22) --
	(475.89,190.22);

\path[draw=drawColor,line width= 0.2pt,line join=round] ( 63.24,271.61) --
	(475.89,271.61);

\path[draw=drawColor,line width= 0.2pt,line join=round] ( 63.24, 68.13) --
	(475.89, 68.13);

\path[draw=drawColor,line width= 0.2pt,line join=round] ( 63.24,149.52) --
	(475.89,149.52);

\path[draw=drawColor,line width= 0.2pt,line join=round] ( 63.24,230.91) --
	(475.89,230.91);

\path[draw=drawColor,line width= 0.2pt,line join=round] ( 63.24,312.31) --
	(475.89,312.31);

\path[draw=drawColor,line width= 0.2pt,line join=round] (140.61, 56.15) --
	(140.61,331.35);

\path[draw=drawColor,line width= 0.2pt,line join=round] (269.57, 56.15) --
	(269.57,331.35);

\path[draw=drawColor,line width= 0.2pt,line join=round] (398.52, 56.15) --
	(398.52,331.35);
\definecolor{drawColor}{gray}{0.20}

\path[draw=drawColor,line width= 0.6pt,line join=round] (140.61, 72.15) -- (140.61, 74.88);

\path[draw=drawColor,line width= 0.6pt,line join=round] (140.61, 69.54) -- (140.61, 68.66);
\definecolor{fillColor}{RGB}{255,255,255}

\path[draw=drawColor,line width= 0.6pt,fill=fillColor] ( 92.26, 72.15) --
	( 92.26, 69.54) --
	(188.97, 69.54) --
	(188.97, 72.15) --
	( 92.26, 72.15) --
	cycle;

\path[draw=drawColor,line width= 1.2pt] ( 92.26, 69.82) -- (188.97, 69.82);
\definecolor{fillColor}{gray}{0.20}

\path[draw=drawColor,line width= 0.4pt,line join=round,line cap=round,fill=fillColor] (269.57,130.89) circle (  2.04);

\path[draw=drawColor,line width= 0.4pt,line join=round,line cap=round,fill=fillColor] (269.57, 91.42) circle (  2.04);

\path[draw=drawColor,line width= 0.4pt,line join=round,line cap=round,fill=fillColor] (269.57,161.58) circle (  2.04);

\path[draw=drawColor,line width= 0.4pt,line join=round,line cap=round,fill=fillColor] (269.57,155.56) circle (  2.04);

\path[draw=drawColor,line width= 0.4pt,line join=round,line cap=round,fill=fillColor] (269.57,165.23) circle (  2.04);

\path[draw=drawColor,line width= 0.4pt,line join=round,line cap=round,fill=fillColor] (269.57, 80.54) circle (  2.04);

\path[draw=drawColor,line width= 0.4pt,line join=round,line cap=round,fill=fillColor] (269.57,136.02) circle (  2.04);

\path[draw=drawColor,line width= 0.4pt,line join=round,line cap=round,fill=fillColor] (269.57,199.34) circle (  2.04);

\path[draw=drawColor,line width= 0.4pt,line join=round,line cap=round,fill=fillColor] (269.57,251.52) circle (  2.04);

\path[draw=drawColor,line width= 0.4pt,line join=round,line cap=round,fill=fillColor] (269.57, 79.09) circle (  2.04);

\path[draw=drawColor,line width= 0.4pt,line join=round,line cap=round,fill=fillColor] (269.57, 89.77) circle (  2.04);

\path[draw=drawColor,line width= 0.4pt,line join=round,line cap=round,fill=fillColor] (269.57, 83.15) circle (  2.04);

\path[draw=drawColor,line width= 0.4pt,line join=round,line cap=round,fill=fillColor] (269.57, 87.17) circle (  2.04);

\path[draw=drawColor,line width= 0.4pt,line join=round,line cap=round,fill=fillColor] (269.57,310.05) circle (  2.04);

\path[draw=drawColor,line width= 0.4pt,line join=round,line cap=round,fill=fillColor] (269.57,203.39) circle (  2.04);

\path[draw=drawColor,line width= 0.6pt,line join=round] (269.57, 74.28) -- (269.57, 77.44);

\path[draw=drawColor,line width= 0.6pt,line join=round] (269.57, 71.45) -- (269.57, 68.69);
\definecolor{fillColor}{RGB}{255,255,255}

\path[draw=drawColor,line width= 0.6pt,fill=fillColor] (221.21, 74.28) --
	(221.21, 71.45) --
	(317.92, 71.45) --
	(317.92, 74.28) --
	(221.21, 74.28) --
	cycle;

\path[draw=drawColor,line width= 1.2pt] (221.21, 72.54) -- (317.92, 72.54);
\definecolor{fillColor}{gray}{0.20}

\path[draw=drawColor,line width= 0.4pt,line join=round,line cap=round,fill=fillColor] (398.52,203.23) circle (  2.04);

\path[draw=drawColor,line width= 0.4pt,line join=round,line cap=round,fill=fillColor] (398.52,220.96) circle (  2.04);

\path[draw=drawColor,line width= 0.4pt,line join=round,line cap=round,fill=fillColor] (398.52,318.84) circle (  2.04);

\path[draw=drawColor,line width= 0.4pt,line join=round,line cap=round,fill=fillColor] (398.52,241.29) circle (  2.04);

\path[draw=drawColor,line width= 0.4pt,line join=round,line cap=round,fill=fillColor] (398.52,143.66) circle (  2.04);

\path[draw=drawColor,line width= 0.4pt,line join=round,line cap=round,fill=fillColor] (398.52,265.57) circle (  2.04);

\path[draw=drawColor,line width= 0.4pt,line join=round,line cap=round,fill=fillColor] (398.52,271.30) circle (  2.04);

\path[draw=drawColor,line width= 0.4pt,line join=round,line cap=round,fill=fillColor] (398.52,256.35) circle (  2.04);

\path[draw=drawColor,line width= 0.4pt,line join=round,line cap=round,fill=fillColor] (398.52,225.81) circle (  2.04);

\path[draw=drawColor,line width= 0.4pt,line join=round,line cap=round,fill=fillColor] (398.52,171.73) circle (  2.04);

\path[draw=drawColor,line width= 0.4pt,line join=round,line cap=round,fill=fillColor] (398.52,211.52) circle (  2.04);

\path[draw=drawColor,line width= 0.4pt,line join=round,line cap=round,fill=fillColor] (398.52,292.61) circle (  2.04);

\path[draw=drawColor,line width= 0.6pt,line join=round] (398.52, 94.56) -- (398.52,108.24);

\path[draw=drawColor,line width= 0.6pt,line join=round] (398.52, 72.81) -- (398.52, 69.26);
\definecolor{fillColor}{RGB}{255,255,255}

\path[draw=drawColor,line width= 0.6pt,fill=fillColor] (350.16, 94.56) --
	(350.16, 72.81) --
	(446.88, 72.81) --
	(446.88, 94.56) --
	(350.16, 94.56) --
	cycle;

\path[draw=drawColor,line width= 1.2pt] (350.16, 76.85) -- (446.88, 76.85);
\end{scope}
\begin{scope}
\definecolor{drawColor}{gray}{0.30}

\node[text=drawColor,anchor=base east,inner sep=0pt, outer sep=0pt] at ( 58.07, 64.17) {0};

\node[text=drawColor,anchor=base east,inner sep=0pt, outer sep=0pt] at ( 58.07,145.56) {100};

\node[text=drawColor,anchor=base east,inner sep=0pt, outer sep=0pt] at ( 58.07,226.95) {200};

\node[text=drawColor,anchor=base east,inner sep=0pt, outer sep=0pt] at ( 58.07,308.35) {300};
\end{scope}
\begin{scope}
\definecolor{drawColor}{gray}{0.30}

\node[text=drawColor,anchor=base,inner sep=0pt, outer sep=0pt] at (140.61, 43.06) {handful};

\node[text=drawColor,anchor=base,inner sep=0pt, outer sep=0pt] at (269.57, 43.06) {few};

\node[text=drawColor,anchor=base,inner sep=0pt, outer sep=0pt] at (398.52, 43.06) {many};
\end{scope}
\end{tikzpicture}\\[0ex]
    \caption{Time per number of permutations.}
    \label{fig:z3-time-per-perms}
  \end{subfigure}
  \quad
  \begin{subfigure}[t]{0.45\textwidth}
    \centering
\input{z3_time_per_vars.tex}
    \caption{Time per number of variables.}
    \label{fig:z3-time-per-vars}
  \end{subfigure}
  \caption{Timings for Z3 experiments (tactic \texttt{smt} disabled) grouped by number of variables or number of permutations. The times are shown in seconds. The graph considers only the instances which did finish within the time bound of \SI{6}{\minute}.}
  \label{fig:z3-time-per-vars-perms_no_tactic_smt}
\end{figure*}

\section{Conclusions}\label{sec:conclusion}
We tested the method for automatically generating polynomial symmetry breakers \cite{erascu2026automaticgenerationpolynomialsymmetry} with
mathematical programming solvers such as Gurobi \cite{gurobi}, CPLEX \cite{manual1987ibm}, and SCIP \cite{hojny2025scip}, the non-
linear and combinatorial optimization solver Hexaly \cite{hexaly}, and the SMT solver Z3 \cite{de2008z3}.

The Gurobi results confirm the conclusions from \cite{erascu2026automaticgenerationpolynomialsymmetry} (see \autoref{sec:gurobi}). The other solvers show more varied behavior. In particular, quadratic constraints are not supported as efficiently as in Gurobi. CPLEX preprocesses the quadratic inequalities, which negatively affects performance; linearizing the breakers using the method from~\cite{balas1964extension} reduced the advantage of quadratic breakers over linear ones (see \autoref{sec:gurobi} and~\ref{sec:cplex}). Thus, quadratic breakers can currently be recommended only when the solver explicitly supports them.

Adding breakers, either quadratic or linear, can also worsen performance on instances that are already solved very quickly without them (see \autoref{sec:gurobi}). This suggests generating quadratic breakers dynamically after a time threshold rather than adding them statically.

The number and size of breakers strongly affect performance, especially in the Z3 satisfiability-only experiments (see \autoref{sec:z3}). Small variable sets (``one'', ``handful'') and a smaller number of breakers (``handful'') avoided time-outs, while other categories had higher time-out rates. Interestingly, ``many'' permutations performed better than ``few'' permutations.


\bibliographystyle{plain}
\bibliography{symmetry}
\appendix
\appendix
\section{Appendix}
\label{app:appendix}

We provide additional details for the breaker generation
process. Similar
to~\cite{erascu2026automaticgenerationpolynomialsymmetry}, breaker
generation relies on templates which are governed by three parameters:
\begin{enumerate*}[label={}, itemjoin={\ }, itemjoin*={\ and\ }, afterlabel={}]
\item the shape of the base polynomial $h$,
\item the number of variables in the base polynomial $h$,
\item the number $\ell$ of permutations $P_1, \ldots, P_\ell$ applied
  to the base polynomial.
\end{enumerate*}
The different possible shapes are summarized in
\autoref{tab:shapes}---they are exactly the same as the ones used
in~\cite{erascu2026automaticgenerationpolynomialsymmetry}. Note that
each $x$ in the shape column of \autoref{tab:shapes} stands for a
linear combination of several $x_{ik}$ with different
$i \leq i, k \leq n$ rather than just for a single
variable. Similarly, $y$ denotes a linear combination of several $y_k$
with different $1 \leq k \leq n$. Moreover, the squared versions $x^2$
and $y^2$ do not mean taking the square of the same linear
combinations but rather they symbolize the product of two different
linear combinations of $x_{ik}$ or $y_k$, respectively. The ``name of
the template'' column refers to the names of the breaker files
generated by our program~\cite{github}. The ``keep only degree~$2$''
column shows whether breakers of degree less than $2$ should be
discarded by the program (and thus not be included in the
output). This is necessary since in the current implementation it is
possible that the quadratic terms in $h(P x)$ and $h(x)$ cancel (\eg,
if $P$ only affects variables in the linear terms) leaving us with a
linear breaker. However, because the purpose of our experiments is to
test explicitly quadratic breakers, we have decided not to use any
linear breaker which are created by this kind of accident.

\begin{table}[h!]
  \centering
  \footnotesize
  \begin{tabular}{lll}
    \toprule
    Name of the Template             & Shape       & Keep Only Degree~$2$ \\
    \midrule
    \texttt{only\_x}                 & $x^2$       & yes                  \\
    \texttt{only\_y}                 & $y^2$       & yes                  \\
    \texttt{mixed}                   & $x\, y$     & yes                  \\
    \texttt{separate}                & $x^2 + y^2$ & yes                  \\
    \texttt{linear\_x}               & $x + y^2$   & yes                  \\
    \texttt{linear\_y}               & $x^2 + y$   & yes                  \\
    \texttt{pure\_linear\_only\_x}   & $x$         & no                   \\
    \texttt{pure\_linear\_only\_y}   & $y$         & no                   \\
    \texttt{pure\_linear\_x\_and\_y} & $x + y$     & no                   \\
    \bottomrule
  \end{tabular}
  \caption{Summary of the templates used for generating the random
    polynomial $h$. Each $x$ and $y$ represents a linear combination of
    several randomly selected $x_{ik}$ and $y_k$, respectively. For
    $x^2$ or $y^2$ two different linear combinations were used for each
    factor. The number of variables in each linear combination is controlled
    by a separate parameter.}
  \label{tab:shapes}
\end{table}

The number of variables $x_{ik}$ or $y_k$ used for each linear
combination depends on the number of bins $n$ of the problem instance
for which the breakers are to be generated. We provide an overview in
\autoref{tab:var-numbers} where we explain the four different settings
used in our templates. These are ``one'', ``handful'', ``few'',
``many'', and ``numerous''. Note that the different shapes request
different numbers of variables: generally, the quadratic terms (\ie,
$x^2$, $y^2$, and $x\,y$) contain fewer distinct variables than the
linear terms (\ie $x$ and $y$). Moreover, since there are $n^2$
different $x_{ik}$ but only $n$ different $y_k$, the terms containing
$y_k$ will usually have fewer variables. Thus,
\autoref{tab:var-numbers} can only provide a range for the number of
variables and not an exact value. See the source code
\texttt{quadratic-breakers.org} in~\cite{github} for the precise
numbers of variables used for each template.

\begin{table}[h!]
  \centering
  \footnotesize
  \begin{subtable}[b]{0.4\textwidth} 
    \begin{tabular}{ll}
      \toprule
      Description & Number \\
      \midrule
      handful     & 10     \\
      few         & 50     \\
      many        & 500    \\
      \bottomrule
    \end{tabular}
    \caption{Number of permutations.}
    \label{tab:perm-numbers}
  \end{subtable}
  \quad
  \begin{subtable}[b]{0.55\textwidth}
    \begin{tabular}{llll}
      \toprule
      Description & $20 \leq n < 99$ & $99 \leq n < 700$ & $700 \leq n$ \\
      \midrule
      one         & 1--3             & 1--3              & ---          \\
      handful     & 3--8             & 3--8              & ---          \\
      few         & 7--18            & 9--18             & 9--18        \\
      many        & 15--149          & 85--500           & 144--400     \\
      numerous    & ---              & ---               & 500--1800    \\
      \bottomrule
    \end{tabular}
    \caption{Number of variables.}
    \label{tab:var-numbers}    
  \end{subtable}  
  \caption{Number of permutations and variables. Note that the exact number of variables depends on the template and also on the number $n$ of bins for the problem instance. Not all template sizes were used for all $n$. The number of permutations provides an upper bound on the number of breakers since we filter out breakers which are trivial ($0 \leq 0$) or which do not have degree~$2$.}
  \label{tab:perm-var-numbers}
\end{table}

Finally, there are three options for the number of permutations used
for breaker generation which are shown in
\autoref{tab:perm-numbers}. These correspond directly to $\ell$ in
\autoref{sec:poly-sb}. Note that not all permutations lead to useful
breakers: in the current implementation it is possible that $h(P x)$
and $h(x)$ cancel each other leaving us with the true but useless
inequality $0 \leq 0$. The program will not output these trivial
breakers. Moreover---as described above---, for the quadratic
templates the program will skip all linear breakers generated
accidentally. In addition, it is also possible that the same
permutation (and hence the same breaker) is generated more than
once. The current implementation will filter out duplicate breakers
but not replace them by newly generated breakers. Thus,
\autoref{tab:perm-numbers} gives only an upper bound of the number of
breakers generated. However, we did find that due to the large number
of possible permutations only very few trivial or duplicate breakers
get generated. Thus, the actual number of breakers is usually very
close the values in \autoref{tab:perm-numbers}.

In \autoref{tab:scip-default-disabled}, we report results obtained with SCIP. We used two options for running it: both presolve and symmetry disabled (column ``Disabled'') and the ``Default'' SCIP setting. For lower $n$, it is not obvious if the usage of symmetry breakers helps or not the solution process. 
It is easy to see that for high $n$ the computations did not finish, hence we decided not to continue with further experiments.

\begin{table}[h!]
\centering
\footnotesize
\renewcommand{\arraystretch}{1.1}
\begin{tabular}{cclcccc}
\toprule
& & & \multicolumn{2}{c}{Default} & \multicolumn{2}{c}{Disabled} \\
\cmidrule(lr){4-5} \cmidrule(lr){6-7}
$n$ & Class & Variant & Time (s) & Opt. & Time (in s) & Optimum \\
\midrule
\multirow{2}{*}{99}
& \multirow{2}{*}{9}
& Baseline
& 6.45 & 50
& 3.32 & 50 \\
&
& FV--FP--LinX
& 10.55 & 50
& 9.05 & 50 \\

\addlinespace

\multirow{2}{*}{100}
& \multirow{2}{*}{7}
& Baseline
& 5.62 & 50
& 3.70 & 50 \\

&
& FV--FP--LinX
& 3.22 & 50
& 7.97 & 50 \\

\addlinespace

\multirow{2}{*}{700}
& \multirow{2}{*}{5}
& Baseline
& Timeout & ---
& Timeout & --- \\

&
& FV--FP--LinX
& Timeout & ---
& Timeout & --- \\

\addlinespace

\multirow{2}{*}{900}
& \multirow{2}{*}{3}
& Baseline
& Timeout & ---
& Timeout & --- \\

&
& FV--FP--LinX
& Timeout & ---
& Timeout & --- \\

\bottomrule
\end{tabular}
\caption{SCIP relative performance with default settings and with presolve and symmetry handling disabled, for \(B=100\). FV--FP--LinX denotes the formulation with few variables, few permutations, and linearized \(x\).}
\label{tab:scip-default-disabled}
\end{table}

This appendix reports additional Z3 satisfiability-only results, focusing on timeout behavior and runtime distributions under the two tactic configurations discussed in \autoref{sec:z3}.

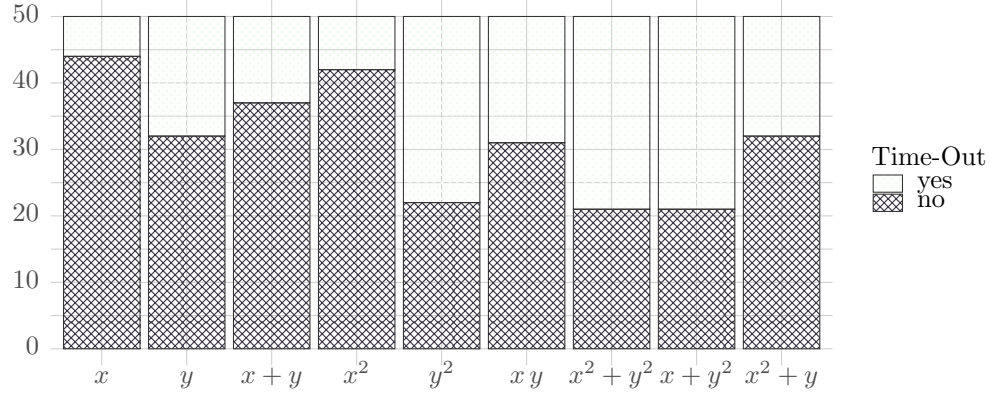
\begin{figure*}[h!]
\centering
\begin{tikzpicture}[x=0.85pt,y=0.5pt]
  \definecolor{fillColor}{RGB}{255,255,255}
\begin{scope}
\definecolor{fillColor}{RGB}{255,255,255}

\end{scope}
\begin{scope}
\definecolor{drawColor}{gray}{0.80}

\path[draw=drawColor,line width= 0.2pt,line join=round] ( 57.50, 93.68) --
	(403.69, 93.68);

\path[draw=drawColor,line width= 0.2pt,line join=round] ( 57.50,143.72) --
	(403.69,143.72);

\path[draw=drawColor,line width= 0.2pt,line join=round] ( 57.50,193.75) --
	(403.69,193.75);

\path[draw=drawColor,line width= 0.2pt,line join=round] ( 57.50,243.79) --
	(403.69,243.79);

\path[draw=drawColor,line width= 0.2pt,line join=round] ( 57.50,293.82) --
	(403.69,293.82);

\path[draw=drawColor,line width= 0.2pt,line join=round] ( 57.50, 68.66) --
	(403.69, 68.66);

\path[draw=drawColor,line width= 0.2pt,line join=round] ( 57.50,118.70) --
	(403.69,118.70);

\path[draw=drawColor,line width= 0.2pt,line join=round] ( 57.50,168.73) --
	(403.69,168.73);

\path[draw=drawColor,line width= 0.2pt,line join=round] ( 57.50,218.77) --
	(403.69,218.77);

\path[draw=drawColor,line width= 0.2pt,line join=round] ( 57.50,268.81) --
	(403.69,268.81);

\path[draw=drawColor,line width= 0.2pt,line join=round] ( 57.50,318.84) --
	(403.69,318.84);

\path[draw=drawColor,line width= 0.2pt,line join=round] ( 80.07, 56.15) --
	( 80.07,331.35);

\path[draw=drawColor,line width= 0.2pt,line join=round] (117.70, 56.15) --
	(117.70,331.35);

\path[draw=drawColor,line width= 0.2pt,line join=round] (155.33, 56.15) --
	(155.33,331.35);

\path[draw=drawColor,line width= 0.2pt,line join=round] (192.96, 56.15) --
	(192.96,331.35);

\path[draw=drawColor,line width= 0.2pt,line join=round] (230.59, 56.15) --
	(230.59,331.35);

\path[draw=drawColor,line width= 0.2pt,line join=round] (268.22, 56.15) --
	(268.22,331.35);

\path[draw=drawColor,line width= 0.2pt,line join=round] (305.85, 56.15) --
	(305.85,331.35);

\path[draw=drawColor,line width= 0.2pt,line join=round] (343.48, 56.15) --
	(343.48,331.35);

\path[draw=drawColor,line width= 0.2pt,line join=round] (381.11, 56.15) --
	(381.11,331.35);
\definecolor{fillColor}{RGB}{127,201,127}

\path[yesstyle] ( 63.14,288.82) rectangle ( 97.01,318.84);
\definecolor{fillColor}{RGB}{190,174,212}

\path[nostyle] ( 63.14, 68.66) rectangle ( 97.01,288.82);
\definecolor{fillColor}{RGB}{127,201,127}

\path[yesstyle] (100.77,228.78) rectangle (134.64,318.84);
\definecolor{fillColor}{RGB}{190,174,212}

\path[nostyle] (100.77, 68.66) rectangle (134.64,228.78);
\definecolor{fillColor}{RGB}{127,201,127}

\path[yesstyle] (138.40,253.79) rectangle (172.27,318.84);
\definecolor{fillColor}{RGB}{190,174,212}

\path[nostyle] (138.40, 68.66) rectangle (172.27,253.79);
\definecolor{fillColor}{RGB}{127,201,127}

\path[yesstyle] (176.03,278.81) rectangle (209.90,318.84);
\definecolor{fillColor}{RGB}{190,174,212}

\path[nostyle] (176.03, 68.66) rectangle (209.90,278.81);
\definecolor{fillColor}{RGB}{127,201,127}

\path[yesstyle] (213.66,178.74) rectangle (247.52,318.84);
\definecolor{fillColor}{RGB}{190,174,212}

\path[nostyle] (213.66, 68.66) rectangle (247.52,178.74);
\definecolor{fillColor}{RGB}{127,201,127}

\path[yesstyle] (251.29,223.77) rectangle (285.15,318.84);
\definecolor{fillColor}{RGB}{190,174,212}

\path[nostyle] (251.29, 68.66) rectangle (285.15,223.77);
\definecolor{fillColor}{RGB}{127,201,127}

\path[yesstyle] (288.92,173.74) rectangle (322.78,318.84);
\definecolor{fillColor}{RGB}{190,174,212}

\path[nostyle] (288.92, 68.66) rectangle (322.78,173.74);
\definecolor{fillColor}{RGB}{127,201,127}

\path[yesstyle] (326.55,173.74) rectangle (360.41,318.84);
\definecolor{fillColor}{RGB}{190,174,212}

\path[nostyle] (326.55, 68.66) rectangle (360.41,173.74);
\definecolor{fillColor}{RGB}{127,201,127}

\path[yesstyle] (364.18,228.78) rectangle (398.04,318.84);
\definecolor{fillColor}{RGB}{190,174,212}

\path[nostyle] (364.18, 68.66) rectangle (398.04,228.78);
\end{scope}
\begin{scope}
\definecolor{drawColor}{gray}{0.30}

\node[text=drawColor,anchor=base east,inner sep=0pt, outer sep=0pt] at ( 52.32, 64.70) {0};

\node[text=drawColor,anchor=base east,inner sep=0pt, outer sep=0pt] at ( 52.32,114.74) {10};

\node[text=drawColor,anchor=base east,inner sep=0pt, outer sep=0pt] at ( 52.32,164.77) {20};

\node[text=drawColor,anchor=base east,inner sep=0pt, outer sep=0pt] at ( 52.32,214.81) {30};

\node[text=drawColor,anchor=base east,inner sep=0pt, outer sep=0pt] at ( 52.32,264.85) {40};

\node[text=drawColor,anchor=base east,inner sep=0pt, outer sep=0pt] at ( 52.32,314.88) {50};
\end{scope}
\begin{scope}
\definecolor{drawColor}{gray}{0.30}

\node[text=drawColor,anchor=base,inner sep=0pt, outer sep=0pt] at ( 80.07, 43.06) {$x$};

\node[text=drawColor,anchor=base,inner sep=0pt, outer sep=0pt] at (117.70, 43.06) {$y$};

\node[text=drawColor,anchor=base,inner sep=0pt, outer sep=0pt] at (155.33, 43.06) {$x + y$};

\node[text=drawColor,anchor=base,inner sep=0pt, outer sep=0pt] at (192.96, 43.06) {$x^2$};

\node[text=drawColor,anchor=base,inner sep=0pt, outer sep=0pt] at (230.59, 43.06) {$y^2$};

\node[text=drawColor,anchor=base,inner sep=0pt, outer sep=0pt] at (268.22, 43.06) {$x\,y$};

\node[text=drawColor,anchor=base,inner sep=0pt, outer sep=0pt] at (305.85, 43.06) {$x^2 + y^2$};

\node[text=drawColor,anchor=base,inner sep=0pt, outer sep=0pt] at (343.48, 43.06) {$x + y^2$};

\node[text=drawColor,anchor=base,inner sep=0pt, outer sep=0pt] at (381.11, 43.06) {$x^2 + y$};
\end{scope}
\begin{scope}
\definecolor{drawColor}{RGB}{0,0,0}

\node[text=drawColor,anchor=base west,inner sep=0pt, outer sep=0pt] at (420.94,207.12) {Time-Out};
\end{scope}
\begin{scope}
\definecolor{fillColor}{RGB}{127,201,127}

\path[yesstyle] (421.68,186.54) rectangle (434.65,199.51);
\end{scope}
\begin{scope}
\definecolor{fillColor}{RGB}{190,174,212}

\path[nostyle] (421.68,172.09) rectangle (434.65,185.06);
\end{scope}
\begin{scope}
\definecolor{drawColor}{RGB}{0,0,0}

\node[text=drawColor,anchor=base west,inner sep=0pt, outer sep=0pt] at (441.14,189.86) {yes};
\end{scope}
\begin{scope}
\definecolor{drawColor}{RGB}{0,0,0}

\node[text=drawColor,anchor=base west,inner sep=0pt, outer sep=0pt] at (441.14,175.40) {no};
\end{scope}
\end{tikzpicture}
\caption{Number of time-outs for the Z3 (tactic \texttt{smt} enabled) experiments by template shape. The graph shows how many instances for each shape did not finish within in the time bound of \SI{10}{\minute} compared to the number of instances which did finish.}
  \label{fig:z3omt-timeouts-per-shape_tactic_smt}
\end{figure*}

\autoref{fig:z3omt-timeouts-per-shape_tactic_smt} shows that the efficiency of the generated symmetry breakers is dependent on the template shape: all templates have both solved instances and time-outs with templates $x$ and $x^2$ giving the fewest time-outs. \autoref{fig:z3omt-finished-per-10sec_tactic_smt} shows that the solved instances are mostly concentrated between roughly $200$ and $350$ seconds.

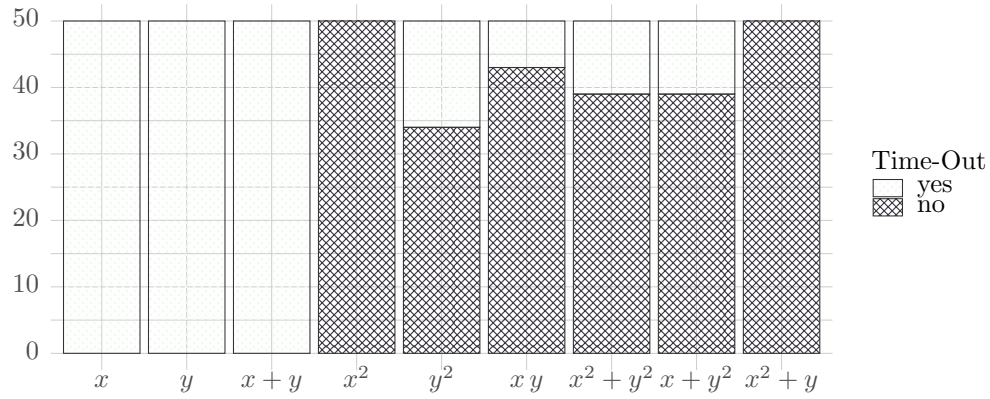
\begin{figure*}[h!]
\centering
\begin{tikzpicture}[x=0.85pt,y=0.5pt]
\definecolor{fillColor}{RGB}{255,255,255}
\begin{scope}
\definecolor{fillColor}{RGB}{255,255,255}

\end{scope}
\begin{scope}
\definecolor{drawColor}{gray}{0.80}

\path[draw=drawColor,line width= 0.2pt,line join=round] ( 57.50, 93.68) --
	(403.69, 93.68);

\path[draw=drawColor,line width= 0.2pt,line join=round] ( 57.50,143.72) --
	(403.69,143.72);

\path[draw=drawColor,line width= 0.2pt,line join=round] ( 57.50,193.75) --
	(403.69,193.75);

\path[draw=drawColor,line width= 0.2pt,line join=round] ( 57.50,243.79) --
	(403.69,243.79);

\path[draw=drawColor,line width= 0.2pt,line join=round] ( 57.50,293.82) --
	(403.69,293.82);

\path[draw=drawColor,line width= 0.2pt,line join=round] ( 57.50, 68.66) --
	(403.69, 68.66);

\path[draw=drawColor,line width= 0.2pt,line join=round] ( 57.50,118.70) --
	(403.69,118.70);

\path[draw=drawColor,line width= 0.2pt,line join=round] ( 57.50,168.73) --
	(403.69,168.73);

\path[draw=drawColor,line width= 0.2pt,line join=round] ( 57.50,218.77) --
	(403.69,218.77);

\path[draw=drawColor,line width= 0.2pt,line join=round] ( 57.50,268.81) --
	(403.69,268.81);

\path[draw=drawColor,line width= 0.2pt,line join=round] ( 57.50,318.84) --
	(403.69,318.84);

\path[draw=drawColor,line width= 0.2pt,line join=round] ( 80.07, 56.15) --
	( 80.07,331.35);

\path[draw=drawColor,line width= 0.2pt,line join=round] (117.70, 56.15) --
	(117.70,331.35);

\path[draw=drawColor,line width= 0.2pt,line join=round] (155.33, 56.15) --
	(155.33,331.35);

\path[draw=drawColor,line width= 0.2pt,line join=round] (192.96, 56.15) --
	(192.96,331.35);

\path[draw=drawColor,line width= 0.2pt,line join=round] (230.59, 56.15) --
	(230.59,331.35);

\path[draw=drawColor,line width= 0.2pt,line join=round] (268.22, 56.15) --
	(268.22,331.35);

\path[draw=drawColor,line width= 0.2pt,line join=round] (305.85, 56.15) --
	(305.85,331.35);

\path[draw=drawColor,line width= 0.2pt,line join=round] (343.48, 56.15) --
	(343.48,331.35);

\path[draw=drawColor,line width= 0.2pt,line join=round] (381.11, 56.15) --
	(381.11,331.35);
\definecolor{fillColor}{RGB}{127,201,127}

\path[yesstyle] ( 63.14, 68.66) rectangle ( 97.01,318.84);

\path[yesstyle] (100.77, 68.66) rectangle (134.64,318.84);

\path[yesstyle] (138.40, 68.66) rectangle (172.27,318.84);
\definecolor{fillColor}{RGB}{190,174,212}

\path[nostyle] (176.03, 68.66) rectangle (209.90,318.84);
\definecolor{fillColor}{RGB}{127,201,127}

\path[yesstyle] (213.66,238.78) rectangle (247.52,318.84);
\definecolor{fillColor}{RGB}{190,174,212}

\path[nostyle] (213.66, 68.66) rectangle (247.52,238.78);
\definecolor{fillColor}{RGB}{127,201,127}

\path[yesstyle] (251.29,283.82) rectangle (285.15,318.84);
\definecolor{fillColor}{RGB}{190,174,212}

\path[nostyle] (251.29, 68.66) rectangle (285.15,283.82);
\definecolor{fillColor}{RGB}{127,201,127}

\path[yesstyle] (288.92,263.80) rectangle (322.78,318.84);
\definecolor{fillColor}{RGB}{190,174,212}

\path[nostyle] (288.92, 68.66) rectangle (322.78,263.80);
\definecolor{fillColor}{RGB}{127,201,127}

\path[yesstyle] (326.55,263.80) rectangle (360.41,318.84);
\definecolor{fillColor}{RGB}{190,174,212}

\path[nostyle] (326.55, 68.66) rectangle (360.41,263.80);

\path[nostyle] (364.18, 68.66) rectangle (398.04,318.84);
\end{scope}
\begin{scope}
\definecolor{drawColor}{gray}{0.30}

\node[text=drawColor,anchor=base east,inner sep=0pt, outer sep=0pt] at ( 52.32, 64.70) {0};

\node[text=drawColor,anchor=base east,inner sep=0pt, outer sep=0pt] at ( 52.32,114.74) {10};

\node[text=drawColor,anchor=base east,inner sep=0pt, outer sep=0pt] at ( 52.32,164.77) {20};

\node[text=drawColor,anchor=base east,inner sep=0pt, outer sep=0pt] at ( 52.32,214.81) {30};

\node[text=drawColor,anchor=base east,inner sep=0pt, outer sep=0pt] at ( 52.32,264.85) {40};

\node[text=drawColor,anchor=base east,inner sep=0pt, outer sep=0pt] at ( 52.32,314.88) {50};
\end{scope}
\begin{scope}
\definecolor{drawColor}{gray}{0.30}

\node[text=drawColor,anchor=base,inner sep=0pt, outer sep=0pt] at ( 80.07, 43.06) {$x$};

\node[text=drawColor,anchor=base,inner sep=0pt, outer sep=0pt] at (117.70, 43.06) {$y$};

\node[text=drawColor,anchor=base,inner sep=0pt, outer sep=0pt] at (155.33, 43.06) {$x + y$};

\node[text=drawColor,anchor=base,inner sep=0pt, outer sep=0pt] at (192.96, 43.06) {$x^2$};

\node[text=drawColor,anchor=base,inner sep=0pt, outer sep=0pt] at (230.59, 43.06) {$y^2$};

\node[text=drawColor,anchor=base,inner sep=0pt, outer sep=0pt] at (268.22, 43.06) {$x\,y$};

\node[text=drawColor,anchor=base,inner sep=0pt, outer sep=0pt] at (305.85, 43.06) {$x^2 + y^2$};

\node[text=drawColor,anchor=base,inner sep=0pt, outer sep=0pt] at (343.48, 43.06) {$x + y^2$};

\node[text=drawColor,anchor=base,inner sep=0pt, outer sep=0pt] at (381.11, 43.06) {$x^2 + y$};
\end{scope}
\begin{scope}
\definecolor{drawColor}{RGB}{0,0,0}

\node[text=drawColor,anchor=base west,inner sep=0pt, outer sep=0pt] at (420.94,207.12) {Time-Out};
\end{scope}
\begin{scope}
\definecolor{fillColor}{RGB}{127,201,127}

\path[yesstyle] (421.68,186.54) rectangle (434.65,199.51);
\end{scope}
\begin{scope}
\definecolor{fillColor}{RGB}{190,174,212}

\path[nostyle] (421.68,172.09) rectangle (434.65,185.06);
\end{scope}
\begin{scope}
\definecolor{drawColor}{RGB}{0,0,0}

\node[text=drawColor,anchor=base west,inner sep=0pt, outer sep=0pt] at (441.14,189.86) {yes};
\end{scope}
\begin{scope}
\definecolor{drawColor}{RGB}{0,0,0}

\node[text=drawColor,anchor=base west,inner sep=0pt, outer sep=0pt] at (441.14,175.40) {no};
\end{scope}
\end{tikzpicture}
\caption{Number of time-outs for the Z3 (tactic \texttt{smt} disabled) experiments by template shape. The graph shows how many instances for each shape did not finish within in the time bound of \SI{6}{\minute} compared to the number of instances which did finish.}
\label{fig:z3-timeouts-per-shape_no_tactic_smt}
\end{figure*}

Across the evaluated configurations, the $x, y, x+y$ instances were consistently associated with the highest computational cost, with all such runs reaching timeout (see \autoref{fig:z3-timeouts-per-shape_no_tactic_smt}), hence they were not included in the subsequent figures. In contrast, the $x^2$ and $x^2+y$ templates exhibited the best and most robust performance, as also reflected in ~\autoref{fig:z3-time-per-shape_no_tactic_smt}.

\begin{figure*}[h!]
\centering
\begin{tikzpicture}[x=0.85pt,y=0.5pt]
\definecolor{fillColor}{RGB}{255,255,255}
\begin{scope}
\definecolor{fillColor}{RGB}{255,255,255}

\end{scope}
\begin{scope}
\definecolor{drawColor}{gray}{0.80}

\path[draw=drawColor,line width= 0.2pt,line join=round] ( 63.24, 98.59) --
	(475.89, 98.59);

\path[draw=drawColor,line width= 0.2pt,line join=round] ( 63.24,158.44) --
	(475.89,158.44);

\path[draw=drawColor,line width= 0.2pt,line join=round] ( 63.24,218.29) --
	(475.89,218.29);

\path[draw=drawColor,line width= 0.2pt,line join=round] ( 63.24,278.14) --
	(475.89,278.14);

\path[draw=drawColor,line width= 0.2pt,line join=round] (142.51, 56.15) --
	(142.51,331.35);

\path[draw=drawColor,line width= 0.2pt,line join=round] (263.52, 56.15) --
	(263.52,331.35);

\path[draw=drawColor,line width= 0.2pt,line join=round] (384.53, 56.15) --
	(384.53,331.35);

\path[draw=drawColor,line width= 0.2pt,line join=round] ( 63.24, 68.66) --
	(475.89, 68.66);

\path[draw=drawColor,line width= 0.2pt,line join=round] ( 63.24,128.51) --
	(475.89,128.51);

\path[draw=drawColor,line width= 0.2pt,line join=round] ( 63.24,188.37) --
	(475.89,188.37);

\path[draw=drawColor,line width= 0.2pt,line join=round] ( 63.24,248.22) --
	(475.89,248.22);

\path[draw=drawColor,line width= 0.2pt,line join=round] ( 63.24,308.07) --
	(475.89,308.07);

\path[draw=drawColor,line width= 0.2pt,line join=round] ( 82.00, 56.15) --
	( 82.00,331.35);

\path[draw=drawColor,line width= 0.2pt,line join=round] (203.01, 56.15) --
	(203.01,331.35);

\path[draw=drawColor,line width= 0.2pt,line join=round] (324.02, 56.15) --
	(324.02,331.35);

\path[draw=drawColor,line width= 0.2pt,line join=round] (445.03, 56.15) --
	(445.03,331.35);
\definecolor{fillColor}{gray}{0.35}

\path[nostyle] ( 82.00, 68.66) rectangle ( 94.10,318.84);

\path[nostyle] ( 94.10, 68.66) rectangle (106.20, 89.01);

\path[nostyle] (106.20, 68.66) rectangle (118.30, 74.65);

\path[nostyle] (118.30, 68.66) rectangle (130.40, 71.06);

\path[nostyle] (130.40, 68.66) rectangle (142.51, 69.86);

\path[nostyle] (142.51, 68.66) rectangle (154.61, 68.66);

\path[nostyle] (154.61, 68.66) rectangle (166.71, 68.66);

\path[nostyle] (166.71, 68.66) rectangle (178.81, 69.86);

\path[nostyle] (178.81, 68.66) rectangle (190.91, 69.86);

\path[nostyle] (190.91, 68.66) rectangle (203.01, 69.86);

\path[nostyle] (203.01, 68.66) rectangle (215.11, 69.86);

\path[nostyle] (215.11, 68.66) rectangle (227.21, 71.06);

\path[nostyle] (227.21, 68.66) rectangle (239.31, 69.86);

\path[nostyle] (239.31, 68.66) rectangle (251.42, 68.66);

\path[nostyle] (251.42, 68.66) rectangle (263.52, 68.66);

\path[nostyle] (263.52, 68.66) rectangle (275.62, 68.66);

\path[nostyle] (275.62, 68.66) rectangle (287.72, 72.25);

\path[nostyle] (287.72, 68.66) rectangle (299.82, 69.86);

\path[nostyle] (299.82, 68.66) rectangle (311.92, 69.86);

\path[nostyle] (311.92, 68.66) rectangle (324.02, 69.86);

\path[nostyle] (324.02, 68.66) rectangle (336.12, 68.66);

\path[nostyle] (336.12, 68.66) rectangle (348.22, 69.86);

\path[nostyle] (348.22, 68.66) rectangle (360.32, 69.86);

\path[nostyle] (360.32, 68.66) rectangle (372.43, 69.86);

\path[nostyle] (372.43, 68.66) rectangle (384.53, 71.06);

\path[nostyle] (384.53, 68.66) rectangle (396.63, 68.66);

\path[nostyle] (396.63, 68.66) rectangle (408.73, 68.66);

\path[nostyle] (408.73, 68.66) rectangle (420.83, 69.86);

\path[nostyle] (420.83, 68.66) rectangle (432.93, 68.66);

\path[nostyle] (432.93, 68.66) rectangle (445.03, 69.86);

\path[nostyle] (445.03, 68.66) rectangle (457.13, 69.86);
\end{scope}
\begin{scope}
\definecolor{drawColor}{gray}{0.30}

\node[text=drawColor,anchor=base east,inner sep=0pt, outer sep=0pt] at ( 58.07, 64.70) {0};

\node[text=drawColor,anchor=base east,inner sep=0pt, outer sep=0pt] at ( 58.07,124.55) {50};

\node[text=drawColor,anchor=base east,inner sep=0pt, outer sep=0pt] at ( 58.07,184.41) {100};

\node[text=drawColor,anchor=base east,inner sep=0pt, outer sep=0pt] at ( 58.07,244.26) {150};

\node[text=drawColor,anchor=base east,inner sep=0pt, outer sep=0pt] at ( 58.07,304.11) {200};
\end{scope}
\begin{scope}
\definecolor{drawColor}{gray}{0.30}

\node[text=drawColor,anchor=base,inner sep=0pt, outer sep=0pt] at ( 82.00, 43.06) {0};

\node[text=drawColor,anchor=base,inner sep=0pt, outer sep=0pt] at (203.01, 43.06) {100};

\node[text=drawColor,anchor=base,inner sep=0pt, outer sep=0pt] at (324.02, 43.06) {200};

\node[text=drawColor,anchor=base,inner sep=0pt, outer sep=0pt] at (445.03, 43.06) {300};
\end{scope}
\end{tikzpicture}
\caption{Z3 (tactic \texttt{smt} disabled) count of how many instances finished within which time
  interval. The intervals have a width of \SI{10}{\second} starting at \SI{0}{\second}. The graph shows only the instances which finished within the time bound of \SI{6}{\minute}. In addition there were 37 time-outs.}
  \label{fig:z3-finished-per-10sec_no_tactic_smt}
\end{figure*}
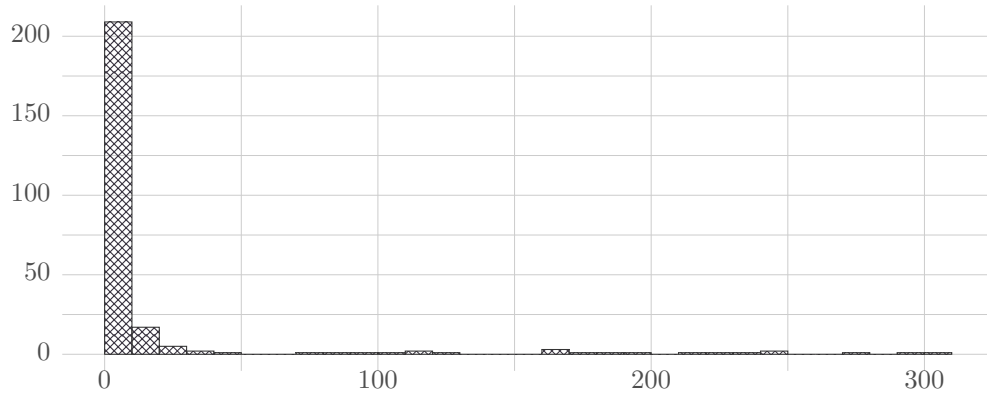

Finally, ~\autoref{fig:z3-finished-per-10sec_no_tactic_smt} shows that most benchmark instances that were solved finished relatively early, within the first 10 seconds, showing that overall the quadratic breakers are beneficial.

\begin{figure*}[h!]
\centering
\begin{tikzpicture}[x=0.85pt, y=0.5pt]
\definecolor{fillColor}{RGB}{255,255,255}
\begin{scope}
\definecolor{fillColor}{RGB}{255,255,255}

\end{scope}
\begin{scope}
\definecolor{drawColor}{gray}{0.80}

\path[draw=drawColor,line width= 0.2pt,line join=round] ( 57.50, 93.68) --
	(475.89, 93.68);

\path[draw=drawColor,line width= 0.2pt,line join=round] ( 57.50,143.72) --
	(475.89,143.72);

\path[draw=drawColor,line width= 0.2pt,line join=round] ( 57.50,193.75) --
	(475.89,193.75);

\path[draw=drawColor,line width= 0.2pt,line join=round] ( 57.50,243.79) --
	(475.89,243.79);

\path[draw=drawColor,line width= 0.2pt,line join=round] ( 57.50,293.82) --
	(475.89,293.82);

\path[draw=drawColor,line width= 0.2pt,line join=round] ( 67.67, 56.15) --
	( 67.67,331.35);

\path[draw=drawColor,line width= 0.2pt,line join=round] (156.12, 56.15) --
	(156.12,331.35);

\path[draw=drawColor,line width= 0.2pt,line join=round] (244.58, 56.15) --
	(244.58,331.35);

\path[draw=drawColor,line width= 0.2pt,line join=round] (333.03, 56.15) --
	(333.03,331.35);

\path[draw=drawColor,line width= 0.2pt,line join=round] (421.49, 56.15) --
	(421.49,331.35);

\path[draw=drawColor,line width= 0.2pt,line join=round] ( 57.50, 68.66) --
	(475.89, 68.66);

\path[draw=drawColor,line width= 0.2pt,line join=round] ( 57.50,118.70) --
	(475.89,118.70);

\path[draw=drawColor,line width= 0.2pt,line join=round] ( 57.50,168.73) --
	(475.89,168.73);

\path[draw=drawColor,line width= 0.2pt,line join=round] ( 57.50,218.77) --
	(475.89,218.77);

\path[draw=drawColor,line width= 0.2pt,line join=round] ( 57.50,268.81) --
	(475.89,268.81);

\path[draw=drawColor,line width= 0.2pt,line join=round] ( 57.50,318.84) --
	(475.89,318.84);

\path[draw=drawColor,line width= 0.2pt,line join=round] (111.90, 56.15) --
	(111.90,331.35);

\path[draw=drawColor,line width= 0.2pt,line join=round] (200.35, 56.15) --
	(200.35,331.35);

\path[draw=drawColor,line width= 0.2pt,line join=round] (288.81, 56.15) --
	(288.81,331.35);

\path[draw=drawColor,line width= 0.2pt,line join=round] (377.26, 56.15) --
	(377.26,331.35);

\path[draw=drawColor,line width= 0.2pt,line join=round] (465.72, 56.15) --
	(465.72,331.35);
\definecolor{fillColor}{gray}{0.35}

\path[nostyle] ( 76.51, 68.66) rectangle ( 85.36, 78.67);

\path[nostyle] ( 85.36, 68.66) rectangle ( 94.20, 78.67);

\path[nostyle] ( 94.20, 68.66) rectangle (103.05,148.72);

\path[nostyle] (103.05, 68.66) rectangle (111.90, 98.68);

\path[nostyle] (111.90, 68.66) rectangle (120.74,118.70);

\path[nostyle] (120.74, 68.66) rectangle (129.59,268.81);

\path[nostyle] (129.59, 68.66) rectangle (138.43,158.73);

\path[nostyle] (138.43, 68.66) rectangle (147.28,268.81);

\path[nostyle] (147.28, 68.66) rectangle (156.12,288.82);

\path[nostyle] (156.12, 68.66) rectangle (164.97,318.84);

\path[nostyle] (164.97, 68.66) rectangle (173.81,228.78);

\path[nostyle] (173.81, 68.66) rectangle (182.66,188.75);

\path[nostyle] (182.66, 68.66) rectangle (191.51,248.79);

\path[nostyle] (191.51, 68.66) rectangle (200.35,118.70);

\path[nostyle] (200.35, 68.66) rectangle (209.20,228.78);

\path[nostyle] (209.20, 68.66) rectangle (218.04, 88.68);

\path[nostyle] (218.04, 68.66) rectangle (226.89,168.73);

\path[nostyle] (226.89, 68.66) rectangle (235.73,198.76);

\path[nostyle] (235.73, 68.66) rectangle (244.58,118.70);

\path[nostyle] (244.58, 68.66) rectangle (253.42,148.72);

\path[nostyle] (253.42, 68.66) rectangle (262.27, 68.66);

\path[nostyle] (262.27, 68.66) rectangle (271.12,108.69);

\path[nostyle] (271.12, 68.66) rectangle (279.96,108.69);

\path[nostyle] (279.96, 68.66) rectangle (288.81,108.69);

\path[nostyle] (288.81, 68.66) rectangle (297.65,128.71);

\path[nostyle] (297.65, 68.66) rectangle (306.50, 98.68);

\path[nostyle] (306.50, 68.66) rectangle (315.34,108.69);

\path[nostyle] (315.34, 68.66) rectangle (324.19, 98.68);

\path[nostyle] (324.19, 68.66) rectangle (333.03,118.70);

\path[nostyle] (333.03, 68.66) rectangle (341.88, 98.68);

\path[nostyle] (341.88, 68.66) rectangle (350.73, 68.66);

\path[nostyle] (350.73, 68.66) rectangle (359.57, 98.68);

\path[nostyle] (359.57, 68.66) rectangle (368.42, 78.67);

\path[nostyle] (368.42, 68.66) rectangle (377.26, 88.68);

\path[nostyle] (377.26, 68.66) rectangle (386.11, 88.68);

\path[nostyle] (386.11, 68.66) rectangle (394.95,128.71);

\path[nostyle] (394.95, 68.66) rectangle (403.80,108.69);

\path[nostyle] (403.80, 68.66) rectangle (412.64, 88.68);

\path[nostyle] (412.64, 68.66) rectangle (421.49, 68.66);

\path[nostyle] (421.49, 68.66) rectangle (430.34, 88.68);

\path[nostyle] (430.34, 68.66) rectangle (439.18, 88.68);

\path[nostyle] (439.18, 68.66) rectangle (448.03, 78.67);

\path[nostyle] (448.03, 68.66) rectangle (456.87, 88.68);
\definecolor{drawColor}{RGB}{255,0,0}

\path[baseline] (282.34, 56.15) -- (282.34,331.35) node[above] {Baseline};
\end{scope}
\begin{scope}
\definecolor{drawColor}{gray}{0.30}

\node[text=drawColor,anchor=base east,inner sep=0pt, outer sep=0pt] at ( 52.32, 64.70) {0};

\node[text=drawColor,anchor=base east,inner sep=0pt, outer sep=0pt] at ( 52.32,114.74) {5};

\node[text=drawColor,anchor=base east,inner sep=0pt, outer sep=0pt] at ( 52.32,164.77) {10};

\node[text=drawColor,anchor=base east,inner sep=0pt, outer sep=0pt] at ( 52.32,214.81) {15};

\node[text=drawColor,anchor=base east,inner sep=0pt, outer sep=0pt] at ( 52.32,264.85) {20};

\node[text=drawColor,anchor=base east,inner sep=0pt, outer sep=0pt] at ( 52.32,314.88) {25};
\end{scope}
\begin{scope}
\definecolor{drawColor}{gray}{0.30}

\node[text=drawColor,anchor=base,inner sep=0pt, outer sep=0pt] at (111.90, 43.06) {200};

\node[text=drawColor,anchor=base,inner sep=0pt, outer sep=0pt] at (200.35, 43.06) {300};

\node[text=drawColor,anchor=base,inner sep=0pt, outer sep=0pt] at (288.81, 43.06) {400};

\node[text=drawColor,anchor=base,inner sep=0pt, outer sep=0pt] at (377.26, 43.06) {500};

\node[text=drawColor,anchor=base,inner sep=0pt, outer sep=0pt] at (465.72, 43.06) {600};
\end{scope}
\end{tikzpicture}
\caption{Z3 (tactic \texttt{smt} enabled) count of how many instances finished within which time
  interval. The intervals have a width of \SI{10}{\second} starting at \SI{0}{\second}. The graph shows only the instances which finished within the time bound of \SI{10}{\minute}. In addition there were $168$ time-outs.}
  \label{fig:z3omt-finished-per-10sec_tactic_smt}
\end{figure*}
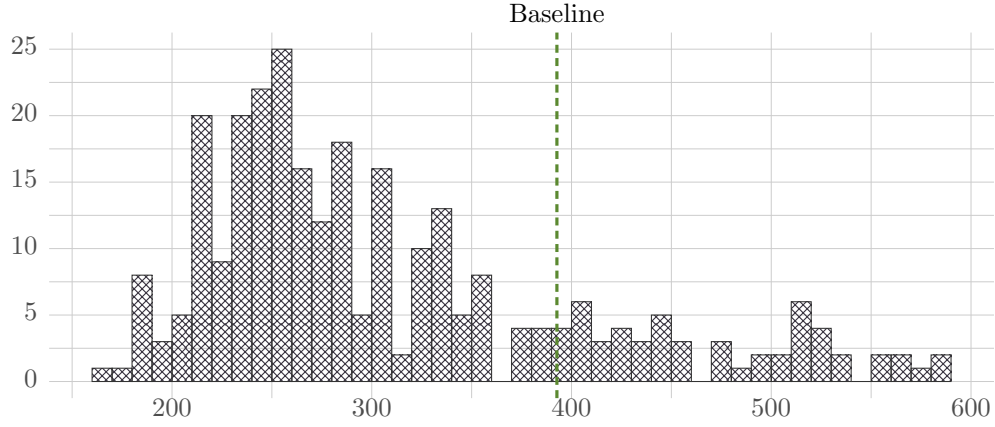

\end{document}